\documentclass[useAMS,usenatbib]{mn2e}

\usepackage{graphicx}
\usepackage{amsmath}
\usepackage{indentfirst}
\def\kms {$\rm km\,s^{-1} \,$}

\title[Nuclear spiral excitation and kinematics in Arp\,102B]{Kinematics and excitation of the nuclear spiral in the active galaxy Arp\,102B}

\author[Guilherme S. Couto et al.]{Guilherme S. Couto$^{1}$\thanks{E-mail:gcouto@if.ufrgs.br}, Thaisa Storchi-Bergmann$^{1}$, David J. Axon$^{2}$, 
\newauthor Andrew Robinson$^{2}$, Preeti Kharb$^{3}$ and Rogemar A. Riffel$^{1,4}$\\
$^{1}$Universidade Federal do Rio Grande do Sul, IF, CP 15051, Porto Alegre 91501-970, RS, Brazil\\
$^{2}$Physics Department, Rochester Institute of Technology, 85 Lomb Memorial Dr., Rochester, NY 14623, USA\\
$^{3}$Indian Institute of Astrophysics, 2nd Block, Koramangala, Bangalore 560034, India\\
$^{4}$Universidade Federal de Santa Maria, Departamento de F\'isica, Centro de Ci\^encias Naturais e Exatas, 97105-900, Santa Maria, RS, Brazil\\}

\begin{document}

\date{Accepted . Received ; in original form }

\pagerange{\pageref{firstpage}--\pageref{lastpage}} \pubyear{}

\maketitle

\label{firstpage}

\begin{abstract}
We present a two-dimensional analysis of the gaseous excitation and kinematics of the inner 2.5\,$\times$\,1.7\,kpc$^2$ of the LINER/Seyfert\,1 galaxy Arp\,102B, from optical spectra obtained with the GMOS integral field spectrograph on the Gemini North telescope at a spatial resolution of $\approx\,250$\,pc. Emission-line flux maps show the same two-armed nuclear spiral we have discovered in previous observations with the HST-ACS camera. One arm reaches 1\,kpc to the east and the other 500\,pc to the west, with a 8.4 GHz VLA bent radio jet correlating with the former. Gas excitation along the arms is low, with line ratios typical of LINERs, and which rule out gas ionization by stars. The gas density is highest ($\approx$\,500 - 900\,cm$^{-3}$) at the nucleus and in the northern border of the east arm, at a region where the radio jet seems to be deflected. Centroid velocity maps suggest that most gas is in rotation in an inclined disk with line of nodes along position angle $\approx 88^{\circ}$, redshifts to the west and blueshifts to the east, with lower blueshifts correlated with the eastern arm and radio jet. This correlation suggests that the jet is interacting with gas in the disk. This interaction is supported by the gas excitation as a function of distance from the nucleus, which requires the contribution from shocks. Channel maps show blueshifts but also some redshifts at the eastern arm and jet location which can be interpreted as originated in the front and back walls of an outflow pushed by the radio jet, suggesting also that the outflow is launched close to the plane of the sky. Principal Component Analysis applied to our data supports this interpretation. We estimate a mass outflow rate along the east arm of $0.26 - 0.32\, \mathrm{M_\odot\, yr^{-1}}$ (depending on the assumed outflow geometry), which is between one and two orders of magnitude higher than the mass accretion rate to the active nucleus, implying that there is mass-loading of the nuclear outflow from circumnuclear gas. The power of this outflow is $0.06 - 0.3 \,\%L_{bol}$. We propose a scenario in which gas has been recently captured by Arp 102B in an interaction with Arp 102A, settling in a disk rotating around the nucleus of Arp 102B and triggering its nuclear activity. A nuclear jet is pushing the circumnuclear gas, giving origin to the nuclear arms. A blueshifted emitting gas knot is observed at $\approx 300$\,pc south-east from the nucleus and can be interpreted as another (more compact) outflow, with a possible counterpart to the north-west.
\end{abstract}


\section{Introduction}

Dusty spirals have been recently discovered to be common nuclear structures (within the inner kpc) of active galaxies \citep{martini03, simoes07}. These compact spirals have been suggested to be possible mechanisms to channel gas inwards to fuel the nuclear supermassive black hole \citep[SMBH,][]{maciejewski04}, with some recent observational work supporting such hypothesis \citep{prieto05, fathi06, storchi07, riffel08, davies09, schnorr11, rifsto11}.

Most nuclear spiral arms seem not to have associated star formation and show low density constrasts, unlike the larger spiral arms in the disks of the galaxies. The dust structures in these arms seem to be signatures of shock fronts of waves propagating in non self-gravitating circumnuclear gas \citep{marpog99}.

In a previous study \citep[][hereafter Paper I]{fathi11}, we have reported the discovery of a two-armed nuclear spiral in a narrow-band H$\alpha$ image of the active galaxy Arp\,102B obtained with the {\it Hubble Space Telescope} (HST) ACS Camera. The origin of the nuclear spiral has been speculated to be the interaction with the companion galaxy Arp\,102A \citep{stauffer83} and could also be tracing inflows triggered by this interaction. An image of the two galaxies\footnote{A picture displaying both Arp 102 galaxies can be viewed in {\it http://www.noao.edu/image\_gallery/html/im0003.html}. A better display of the tidal arms can be seen in {\it http://www.spacebanter.com/attachment.php?attachmentid =4120\&stc=1}} does indeed show a tidal arm originating in the gas rich spiral galaxy Arp102A which connects with Arp102B from the north-east. Nevertheless, in Paper I, we also found that the spiral arm to the east (the other is to the west) is spatially correlated with a bent radio jet, suggesting that the 
spiral could be the result of the interaction of a radio jet with circumnuclear gas at the galaxy. The first detection of radio emission was made by \citet{biermann81}, but it was found to have a jet-like structure only later by \citet{puschell86}. These authors were not able to determine the radio emission mechanism. \citet{caccianiga01}, studying the pair Arp 102A and Arp 102B, demonstrated that the radio jet was originated in an AGN, rather than in a starburst, since its brightness temperature ($10^6 - 10^8$\,K) is too high to be due to starburst emission, whose values are at most $10^5$\,K. In Paper I, we concluded that, in order to gauge the jet impact on the circumnuclear gas it would be necessary to map the gas excitation and kinematics. Such a study would also provide constraints on the nature of the ionizing source as well as on the presence of outflows and/or inflows along the spiral arms, allowing us to investigate their nature. This is the goal of the present study.

One particular characteristic of the active nucleus of Arp 102B is the presence of very broad ($\approx$\,10,000\,kms) and double-peaked Balmer lines in its optical spectra. In fact, Arp 102B is considered to be the prototypical ``double-peaker'' \citep{chen89_1, chen89_2, newman97, floera08}. The Arp 102B nucleus is classified as a LINER/Seyfert 1 being more luminous but other than that similar to the one in NGC 1097 \citep{storchi93, storchi03, fathi06}. Arp 102B is classified as an E0 galaxy, thus the presence of the gaseous spiral arms was unexpected. We adopt a distance to Arp\,102B of 104.9\,Mpc, which gives a scale of 490.2\,pc\,arcsec$^{-1}$ at the galaxy \citep{eracleous04}.

This paper is organized as follows: in Sec. \ref{obs} we describe the observations and data reduction, in Sec. \ref{res} we present the results for the gas excitation and kinematics, in Sec. \ref{disc} we discuss our results and in Sec. \ref{conc} we present our conclusions.

\section{Observations and Data Reduction}
\label{obs}

Two-dimensional optical spectroscopy data were obtained at the Gemini North Telescope with the {\it Gemini Multi-Object Spectrograph Integral Field Unit} (GMOS IFU) \citep{allington02} in April 06, 2007, as part of program GN-2007A-Q-57, and comprise six individual exposures of 900\,s  centred at  $\lambda$\,5850\AA\ with a spectral coverage from $\lambda$4400\AA\ to $\lambda$7300\AA. The B600+\_G5303 grating with the IFU-R mask was used. The spectral resolution is 1.8\AA\ at H\,$\alpha$ ($\approx$\,85 \kms) -- derived from the full width at half maximum (FWHM) of the CuAr lamp emission lines. A similar value is obtained from the FWHM of the night sky emission lines. The angular resolution is 0.6 arcsec (corresponding to 245\,pc at the galaxy) -- adopted as the FWHM of the spatial profile of the flux calibration star, Feige 66. The GMOS IFU has a rectangular field of view, of approximately 5$\farcs$1\,$\times$\,3$\farcs$4, corresponding to 2.5\,kpc$\times$\,1.7\,kpc at the galaxy. The major axis of the IFU 
was oriented along position angle PA=65$^\circ$.

The data reduction was accomplished using tasks in the {\sc gemini.gmos iraf} package as well as generic {\sc iraf} tasks. The reduction procedure included trimming of the images, bias subtraction, flat-fielding, wavelength calibration, s-distortion correction, sky subtraction and relative flux calibration. In order to obtain an absolute flux calibration, we have scaled our IFU data to long-slit data obtained with the $HST$-STIS spectrograph. This was accomplished by dividing our fluxes by 15, in order to match the fluxes obtained with STIS within the same apertures. A correction for telluric absorption lines was also applied, since it affected the sulfur emission lines. In order to improve the spatial resolution, we have corrected the data for differential atmospheric refraction \citep{steiner09} and then applied a Richardson-Lucy deconvolution algorithm \citep{richardson72,lucy74} to the final datacube, which resulted in a spatial resolution of $\approx 0\farcs5$.

\section{Results}
\label{res}

\begin{figure*}
\centering
\includegraphics[width=\textwidth]{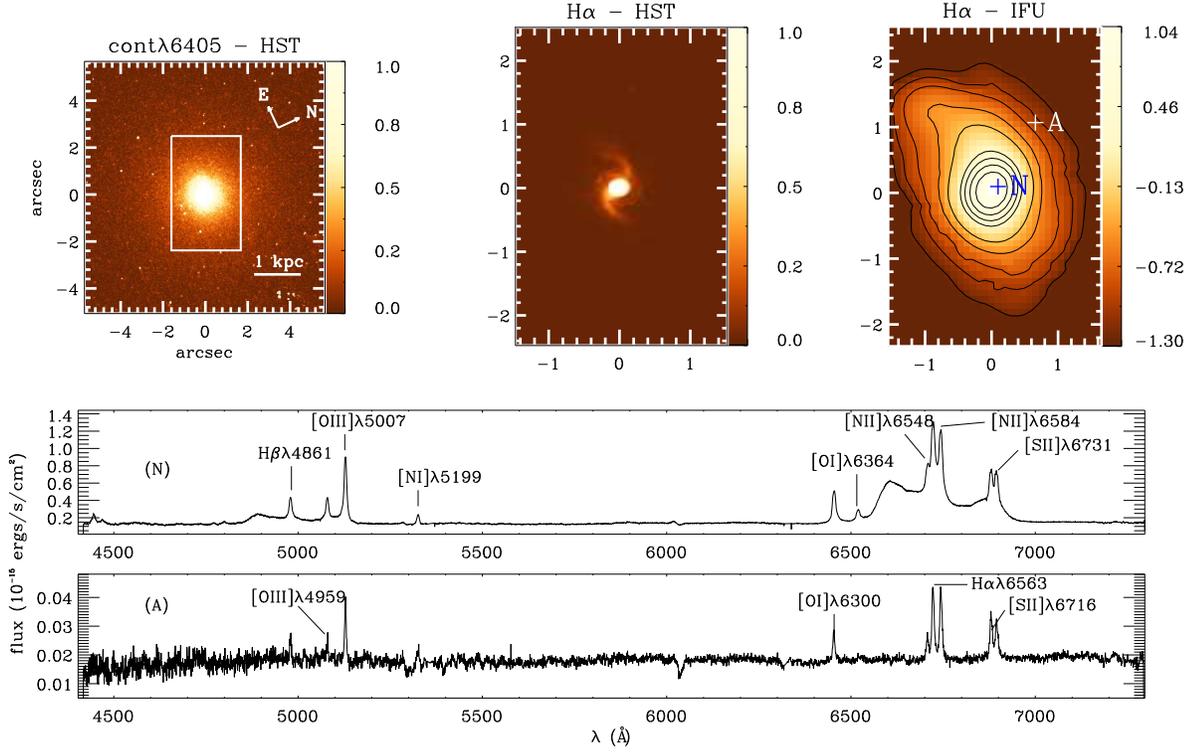}
\caption{Top left: $\lambda$6405$\AA$ continuum map obtained with the HST ACS camera. The rectangle shows the GMOS IFU field-of-view (FOV). Top center: H$\alpha$ flux map obtained with ACS within the IFU FOV. Top right: H$\alpha$ flux map from the IFU spectroscopy. Bottom: Spectra extracted at the positions N (nucleus) and A marked in the top right panel.}
\label{large}
\end{figure*}

The IFU field-of-view and data are illustrated in Fig. \ref{large}, together with the continuum ($\lambda6405$\AA) and  H$\alpha$ flux maps of the inner 26\arcsec\,$\times$\,29\arcsec of Arp\,102B from Paper I,  obtained with the $HST$-ACS camera. In the top left panel, the rectangle shows the IFU field-of-view (FOV) oriented with its longest axis at PA=65$^\circ$. In the top central panel, the HST-ACS continuum-subtracted H$\alpha$ flux map shows the two spiral arms extending to $\sim$500\,pc from the nucleus. These arms are oriented approximately to the east and west of the nucleus and we hereafter refer to them as the ``east arm" and the ``west arm". In the top right panel we show the H$\alpha$ flux map obtained from the IFU spectroscopy. Although showing poorer spatial resolution, this map suggests that the spiral arms extend farther than seen in the $HST$ image, up to $\approx$\,1\,kpc from the nucleus. In the bottom panels we show IFU spectra extracted at the nucleus (position N) and at $\approx$\,1\,
arcsec north-east (position A), with both positions shown in the top right panel. We define the position of the nucleus as the one corresponding to the peak of the continuum emission. The nuclear spectrum shows the known broad double-peaked H$\alpha$ and H$\beta$ profiles, as well as the narrow emission lines H$\beta\lambda$4861, [O\,{\sc iii}]$\lambda\lambda$4959,5007, [N\,{\sc i}]$\lambda$5199, [O\,{\sc i}]$\lambda\lambda$6300,64, [N\,{\sc ii}]$\lambda\lambda$6548,84, H$\alpha\lambda$6563, [S\,{\sc ii}]$\lambda\lambda$6717,31.  All these lines -- with the exception of the broad double-peaked components -- are also present in the extranuclear spectrum, where they are narrower. 

\begin{figure*}
\centering
\includegraphics[width=0.8\textwidth]{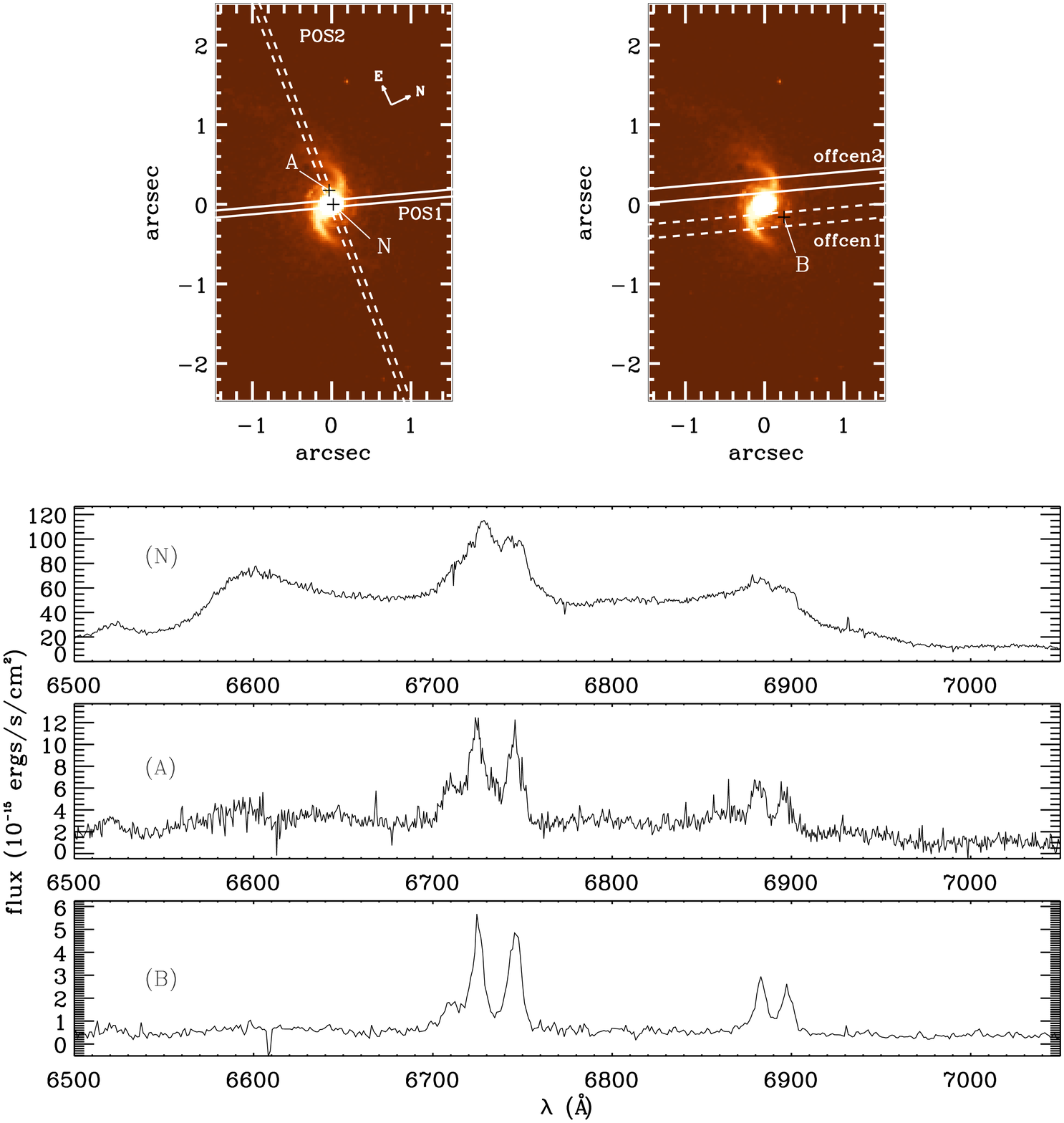}
\caption{Top panels: continuum-subtracted $HST$-ACS H$\alpha$ flux maps showing a representation of the $HST$-STIS slits, from Paper I. Bottom panels: Spectra extracted at positions N, A and B marked in the top panels. The flux maps have the same orientation as Fig. \ref{large}.}
\label{stis_large}
\end{figure*}

In Fig. \ref{stis_large} we present STIS observations from Paper I, for the purpose of comparison with the GMOS-IFU data. In the top panels we show the HST H$\alpha$ image within the same FOV of the IFU observations and the same orientation as Fig. \ref{large}. The STIS slits are shown along the position angles $160.1\,^{\circ}$ for POS1, offcen1 and offcen2 and $85.7\,^{\circ}$ for POS2. Slits identified as POS1 and POS2 cross the nucleus, while offcen1 and offcen2 are 0$\farcs$2 offset from the nucleus. For the measurements we have extracted spectra pixel by pixel, which correspond to $0\farcs05$ for the slits crossing the nucleus and to $0\farcs1$ for the off-center slits. Thus along the slits crossing the nucleus the extractions correspond to apertures of $0\farcs1 \times 0\farcs05$, while for the off-center extractions the apertures are of $0\farcs2 \times 0\farcs1$. In the bottom panel of Fig. \ref{stis_large} we show $HST$-STIS spectra extracted at the nucleus (position N), at $0\farcs1$ east (position A) and at $0\farcs2$ north-west from the center of the offcen2 slit (position B). The nuclear spectrum shows the double-peaked H$\alpha$ and the narrow emission lines [N\,{\sc ii}]$\lambda\lambda$6548,84, H$\alpha\lambda$6563, [S\,{\sc ii}]$\lambda\lambda$6717,31 and a weak [O\,{\sc i}]$\lambda$6364, while the extranuclear ones show only the narrow lines.

\subsection{Emission-line flux distributions}
\label{fluxd}

In order to obtain the gas emission-line fluxes and kinematics at the nucleus and surrounding regions, we had to subtract the broad-line emission from the spectra. Although the double-peaked emitting region is not resolved by our observations, the smearing of the flux due to the seeing makes the double-peaked emission contaminate the extranuclear spectra up to $\approx$\,0$\farcs$3 from the nucleus. Using the programming language IDL\footnote{Interactive Data Language, http://ittvis.com/idl},  we have fitted three Gaussians to the H$\alpha$ broad-line profile, which were then subtracted from the spectra, a procedure which efficiently eliminated the broad component. 

Once we had spectra with no significant broad component, the emission-line flux distributions were obtained by fitting Gaussians to the narrow lines and integrating the corresponding flux after subtraction of the continuum contribution. In the case of the blended lines H\,$\alpha +$ [N\,{\sc ii}], the width of the [N\,{\sc ii}]$\lambda$6548\AA\ and $\lambda$6584\AA\ emission lines were constrained to the same value. The [S\,{\sc ii}]$\lambda$6717\AA\ and $\lambda$6731\AA\ emission-line widths were also constrained to the same value. The value of the [N\,{\sc ii}]$\lambda$6584\AA/$\lambda$6548\AA\ flux ratio recovered from the fits was $\approx$\,3 throughout the FOV, as expected from the nebular physics \citep{ostfer06}.

\begin{figure*}
\centering
\includegraphics[width=0.8\textwidth]{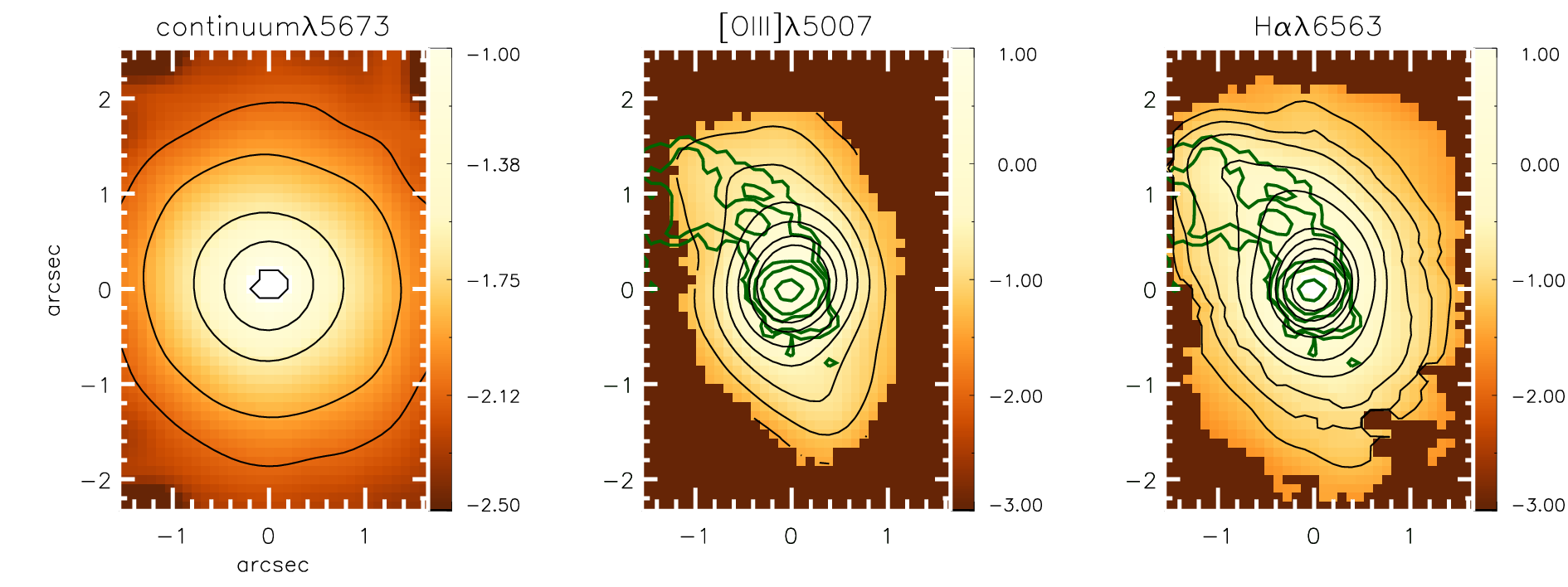}
\caption{Left: Flux map in the continuum. Center: Flux map in [O\,{\sc iii}]$\lambda$5007. Right: Flux map in H$\alpha$. Green contours are from a 8.4 GHz radio image. Flux units are 10$^{-15}$ erg cm$^{-2}$ s$^{-1}$ spaxel$^{-1}$ and are shown in a logarithmic scale.}
\label{flux}
\end{figure*}

Flux maps were obtained in all the emission lines displayed in the spectra of Fig. \ref{large}. [O\,{\sc iii}]$\lambda$5007 and H$\alpha$ flux maps are shown in Fig. \ref{flux}, together with a continuum map at $\lambda$5673, integrated within a spectral window of 100\,\AA. Error maps obtained through Monte Carlo simulations (see Sec. \ref{error}) were used to mask out the emission-line flux distribution for error values in each line $\epsilon_{F}$ relative to the corresponding flux values $F$ larger than $\frac{\epsilon_{F}}{F} = 0.5$. The continuum map shows little structure, with an approximately circularly symmetric flux distribution. The flux distributions in the emission lines are elongated to the east and west, showing the spiral arms first observed in the HST ACS images. The east arm is better defined and extends farther from the nucleus than the west arm for all emission lines except [O\,{\sc iii}], for which the two arms show similar extent ($\approx\,$1 kpc). The green contours are from a 8.4\,GHz 
VLA radio image \citep{fathi11}. Note that the radio structure follows the east spiral arm, suggesting a relation between the emitting gas and the radio structure.

\subsection{Error estimates}
\label{error}

In order to calculate the errors in the measurements of the line fluxes ($\epsilon_{F}$), centroid velocities ($\epsilon_{v}$) and widths ($\epsilon_{\sigma}$), we applied Monte Carlo simulations to the data, adding random noise with a gaussian probability distribution, moduled by the data noise. One hundred realizations were performed for each emission line and the errors were calculated as the standard deviations of the distribution of the parameter values recovered from the one hundred realizations. For each emission line, error maps obtained from Monte Carlo simulations (see Sec. \ref{fluxd}) were used to mask out the flux distribution, by excluding pixels for which $\frac{\epsilon_{F}}{F} > 0.5$, where $\epsilon_{F}$ is the flux error and $F$ the line flux. The same was done for the centroid velocity and velocity dispersion maps (Sec. \ref{ifukin}), for values larger than $\epsilon_v = \epsilon_\sigma = 20$\,\kms.

\subsection{Emission-line ratio maps}

\begin{figure*}
\centering
\includegraphics[width=\textwidth]{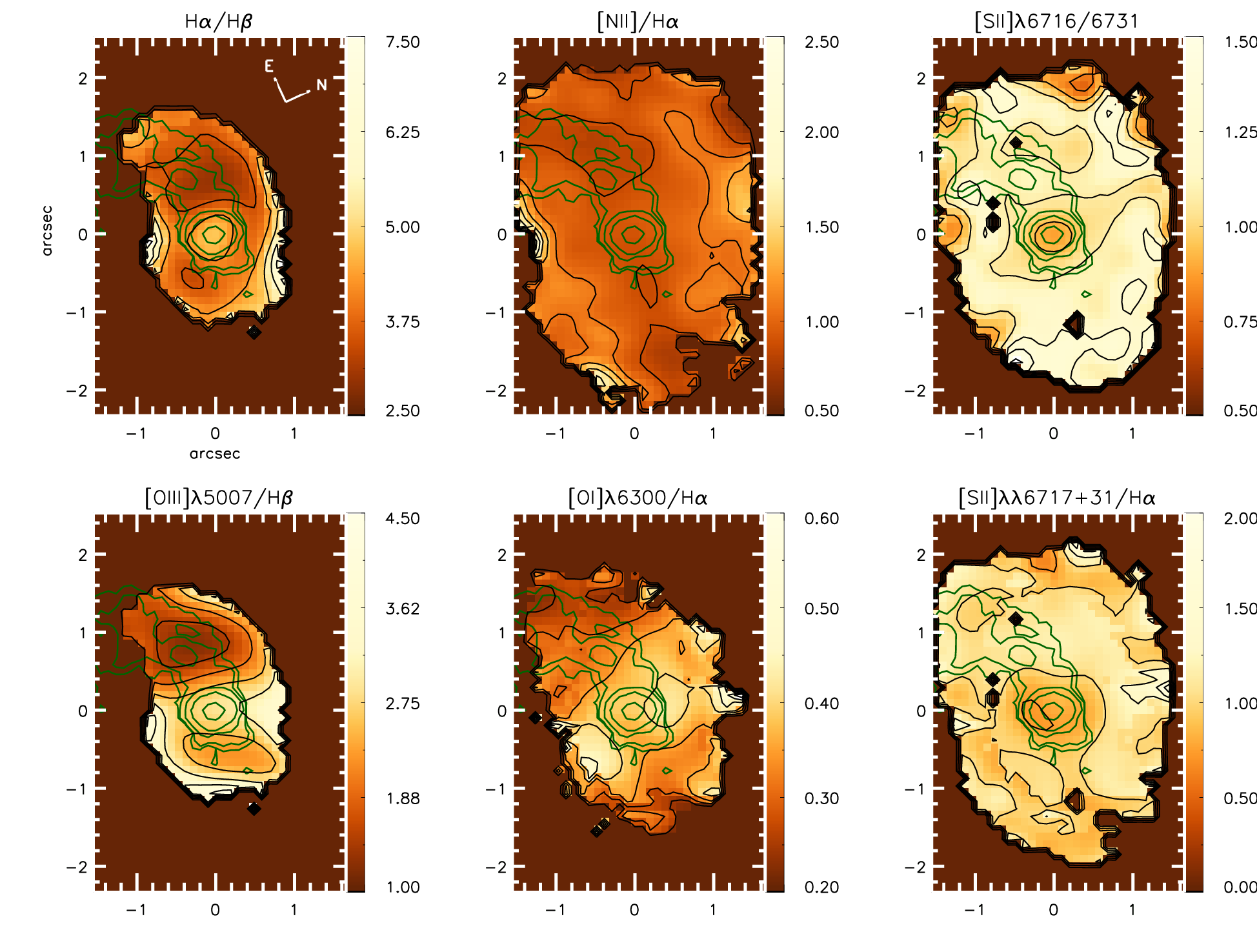}
\caption{Line ratio maps with the green contours showing the radio structure. Top left: H$\alpha$/H$\beta$ ratio. Top middle: [N\,{\sc ii}]6584$\lambda$/H$\alpha$ ratio. Top right: [SII]$\lambda\lambda$6716/6731 emission-line ratio. Bottom left: [O\,{\sc iii}]$\lambda$5007/H$\beta$ ratio. Bottom middle: [O\,{\sc i}]$\lambda$6300/H$\alpha$ ratio, Bottom left: [S\,{\sc ii}]$\lambda\lambda$6717$+$31/H$\alpha$ ratio.}
\label{razao}
\end{figure*}

We have used the emission-line flux distributions to obtain the line ratio maps shown in Fig. \ref{razao}. In the top left panel we show the distribution of  H$\alpha$/H$\beta$ ratios, with the highest values at the nucleus (up to $\approx\,5$) and lowest values ($\approx\,3$) in regions away from the nucleus, mainly along the radio jet. The top middle panel shows that along the central part of the spiral arms, the ratio [N\,{\sc ii}]6584/H$\alpha$ is almost constant at $\approx\,1$, increasing to larger values ($\approx\,1.5$) towards the borders of the arms. The top right panel shows the [S\,{\sc ii}]6717/6731 emission line ratio, which shows the lowest values of $\approx\,$0.9 at the nucleus and highest values ($\approx\,$1.5) towards the west. The [O\,{\sc iii}]5007/H$\beta$ ratio (bottom left panel of Fig. \ref{razao}) shows the highest values of $\geq\,4$ at the nucleus, along a narrow lane running approximately perpendicular to the spiral arms and at the most distant part of the west arm. The lowest values of $\approx\,1.3$ are observed at $\approx$\,1\arcsec\, to the east, in the ``middle'' of the east spiral arm, while to the west, at $\approx\,0\farcs$5 from the nucleus there is a small region with values $\approx\,2$. The [O\,{\sc i}]6300/H$\alpha$ ratio, shown in the bottom middle panel presents the highest values around the nucleus, mostly to north and south, with values $\approx\,0.5$, and lowest values ($\approx\,0.3$) towards distant areas from the nucleus, to the east and west approximately at the locations of the spiral arms. The bottom right panel shows the ratio between the sum of the [S\,{\sc ii}] emission lines and H$\alpha$. The lowest values ($\approx\,0.5$) are seen mainly at the nucleus and to the south, while the highest values, $\approx\,1.2$, are found at the locations where the [O\,{\sc iii}]5007/H$\beta$ ratio is lowest, along the east spiral arm and at the beginning of the west spiral arm. 

\subsection{Gas kinematics}

\subsubsection{Centroid velocities and velocity dispersions - IFU}
\label{ifukin}

Gas velocities were obtained from the centroid wavelengths of the Gaussian curves fitted to the emission-line profiles while velocity dispersions $\sigma$ were obtained from the full-width at half maximum (FWHM) of the Gaussians: $\sigma=$FWHM/2.355. The $\sigma$ values were corrected for the instrumental broadening.

\begin{figure*}
\centering
\includegraphics[scale=0.5]{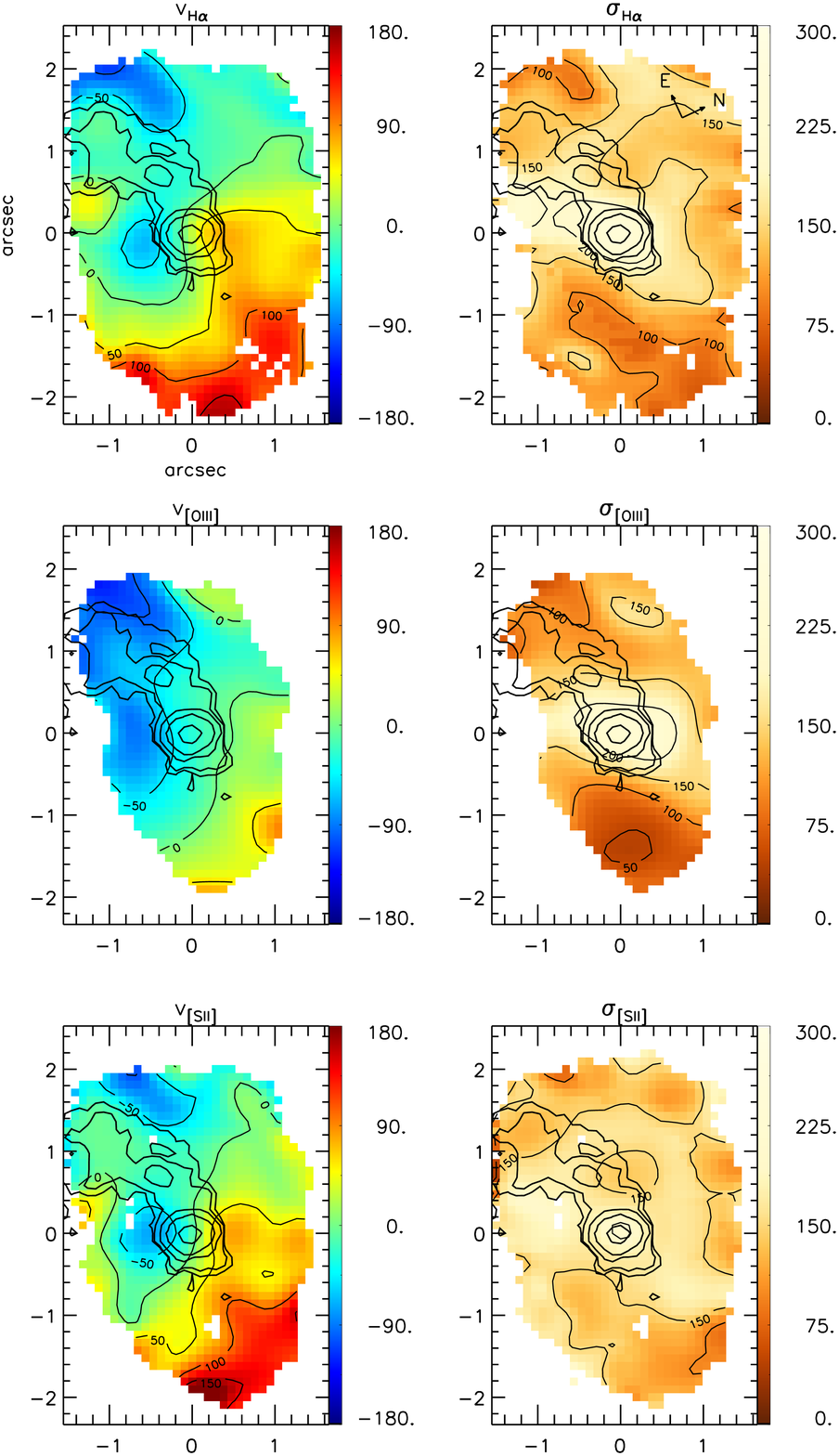}
\caption{Centroid velocity (left panel) and velocity dispersion (right panel) maps for the emission-lines. Units are km s$^{-1}$.}
\label{velocidade}
\end{figure*}

In Fig. \ref{velocidade} we present the centroid velocity maps (left panels) and velocity dispersion maps (right panels) for the gas emission in H\,$\alpha$, [O\,{\sc iii}] and [S\,{\sc ii}]. The velocity values shown in Fig. \ref{velocidade} have had the systemic velocity $v_{sys} = 7213 \pm 35$ \kms of the galaxy subtracted. This value was calculated as the average of the centroid velocities of the H$\beta$, H$\alpha$ and [N\,{\sc ii}]\,$\lambda$\,6584 emitting gas in the 3\,$\times$\,3 central pixels. The [N\,{\sc ii}] velocity field is very similar to that of H$\alpha$ and is thus not shown. The centroid velocity maps (left panels of Fig. \ref{velocidade}) show redshifts to the west and blueshifts to the east. The zero velocity contours show an s-shape with blueshifts to the south-east (left in the figure) and redshifts to the north-west (right in the figure). 

The highest velocity dispersions ($\sigma\,\approx$\,270 km s$^{-1}$) are observed around the nucleus and both to the south-east and north-west (left and right of the nucleus, respectively, in Fig. \ref{velocidade}) for H$\alpha$ and [O\,{\sc iii}], while lower values, down to $\sim$\,100 km s$^{-1}$ are observed elsewhere. The [S\,{\sc ii}] $\sigma$ values are somewhat lower than those of H$\alpha$ and [O\,{\sc iii}], reaching $\approx$ 200 \kms, at the same approximate locations of maximum $\sigma$-values for these lines.

The black contours in Fig. \ref{velocidade} show that the radio structure is mostly associated with blueshifted emission, although at the exact location of the radio emission proper, the blueshifts seem to decrease (by $\approx\,$ 50 \kms) relative to the surrounding regions.

\subsubsection{Channel maps}

\begin{figure*}
\centering
\includegraphics[scale=0.5]{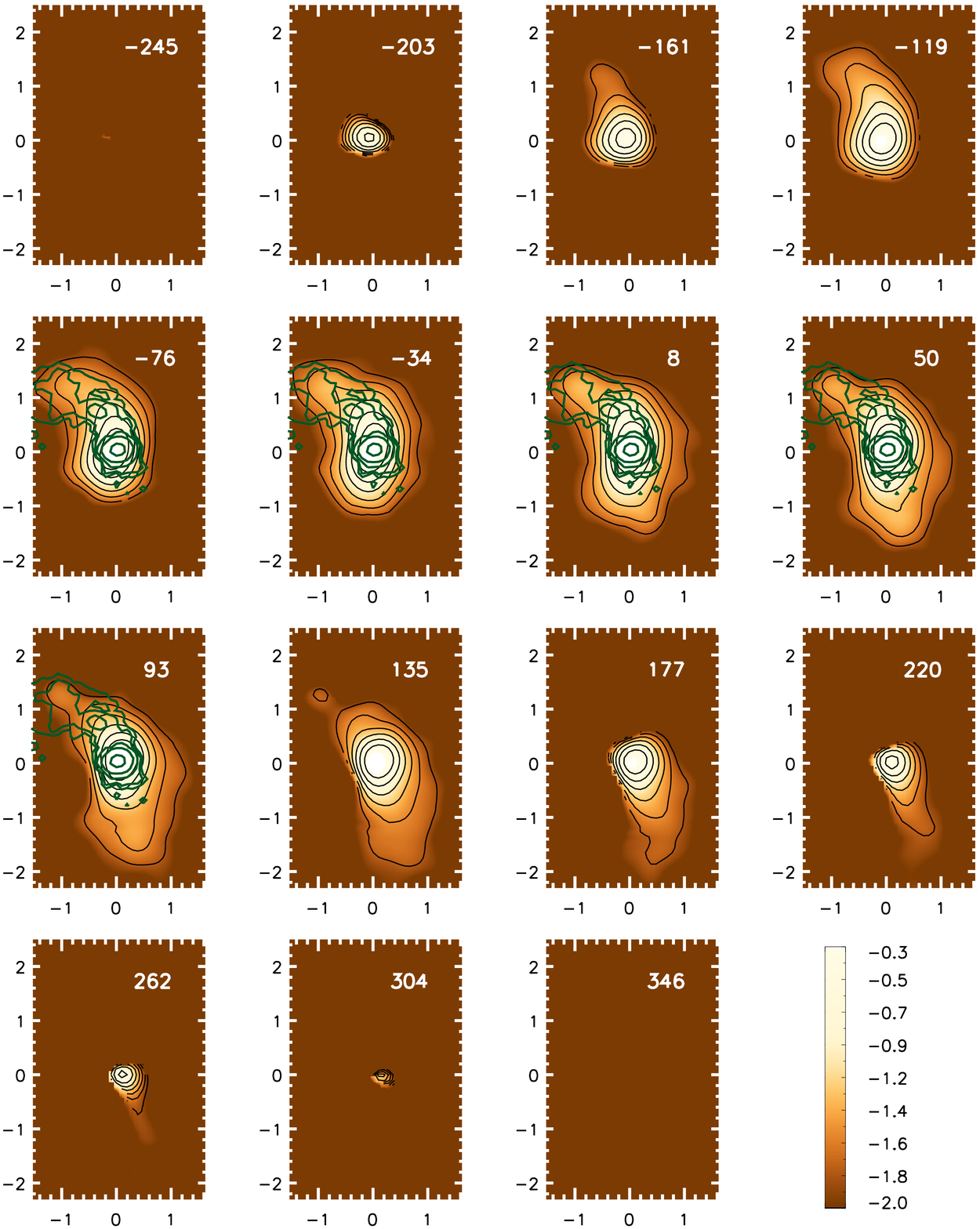}
\caption{Channel maps along the H$\alpha$ gas emission-line profile, in order of increasing velocities shown in the top right of each panel in units of km s$^{-1}$. Flux units are 10$^{-15}$ erg cm$^{-2}$ s$^{-1}$ spaxel$^{-1}$ and are shown in a logarithmic scale.}
\label{cmHalfa}
\end{figure*}

\begin{figure*}
\centering
\includegraphics[scale=0.5]{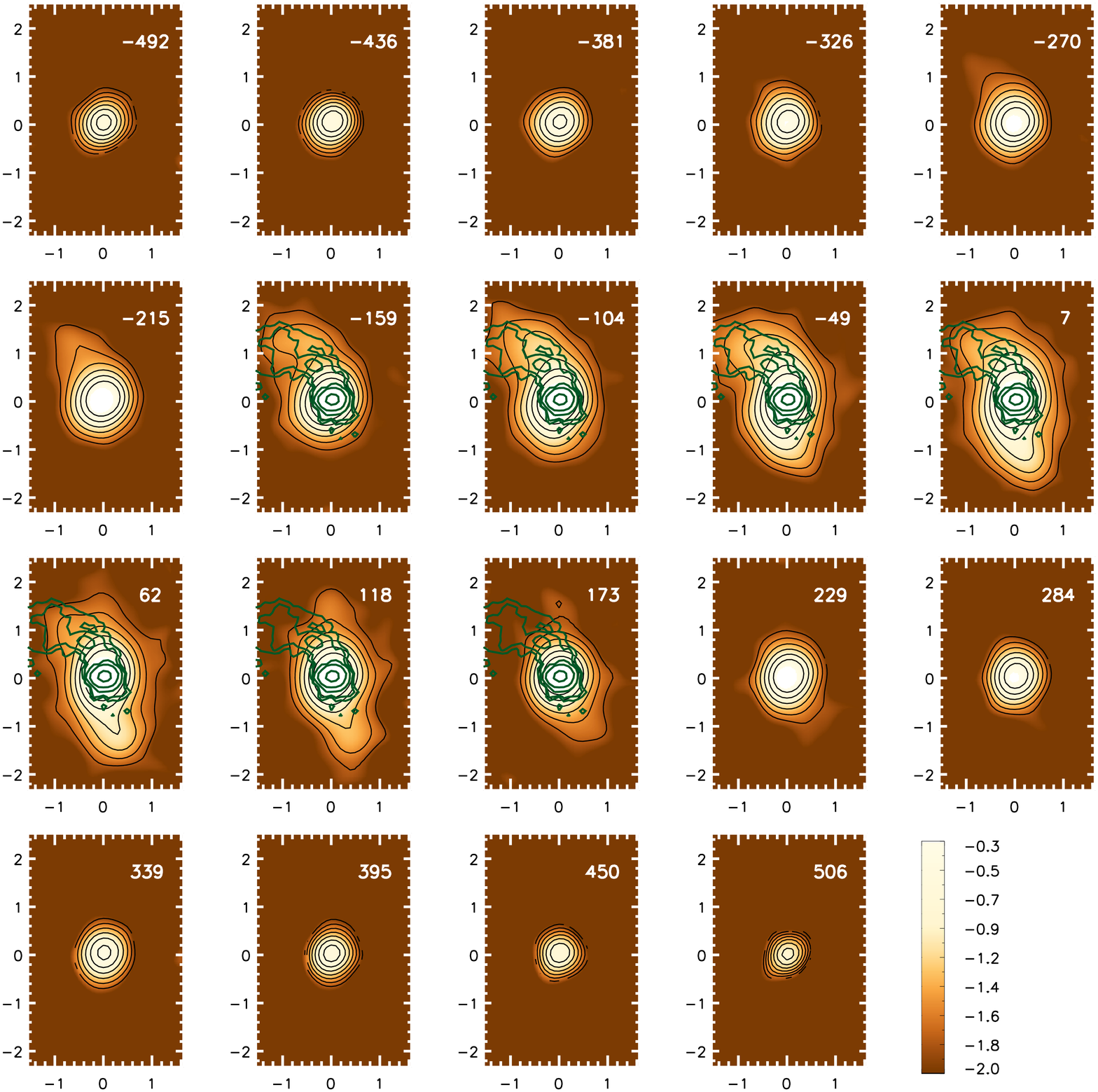}
\caption{Channel maps along the [O\,{\sc iii}]$\lambda$5007 gas emission-line profile, in order of increasing velocities shown in the top right of each panel in units of km s$^{-1}$. Flux units are 10$^{-15}$ erg cm$^{-2}$ s$^{-1}$ spaxel$^{-1}$ and are shown in a logarithmic scale.}
\label{cmOIII}
\end{figure*}

We have mapped the gas kinematics using also channel maps extracted along the emission-line profiles. In Figs. \ref{cmHalfa} and \ref{cmOIII} we show a sequence of these maps extracted within velocity bins of $\sim$\,42\,km\,s$^{-1}$ and $\sim$\,55\,km\,s$^{-1}$ along the H$\alpha$ and [O\,{\sc iii}] emission-line profiles, respectively. They both show the highest blueshifts and redshifts at the nucleus, probably just a consequence of the high velocity dispersions there. 

In the H$\alpha$ channel maps (Fig. \ref{cmHalfa}), the highest blueshifts outside the nucleus are observed along the east spiral arm, extending to 1$\farcs$4 (686 pc), with velocities reaching $-$161 \kms. Along the west spiral arm the highest redshifts are observed at similar distances from the nucleus with velocities of up to 262 \kms. Even though mostly blueshifts are observed along the east spiral arm, there are also some redshifts there, reaching velocities as high as 93 \kms.

The [O\,{\sc iii}] channel maps (Fig. \ref{cmOIII}) present also high blueshifts along the east spiral arm, with velocities reaching $-$270 \kms. Velocities of up to 173 \kms (the highest redshift outside the nucleus) are observed along the west spiral arm. As observed for H$\alpha$, the [O\,{\sc iii}] channel maps feature redshifts as well (as high as 62 \kms) along the east spiral arm. 

The green contours overlaid on the channel maps are from the radio image, supporting a correlation between the radio and the blueshifted gas emission.

\section{Discussion}
\label{disc}

\subsection{Flux distributions}

Flux maps in all emission lines (Fig. \ref{flux}) show the same two-armed spiral observed in the HST image of Paper I, but with the east arm being more clearly observed and extending farther from the nucleus -- to at least 1\,kpc from it. The west spiral arm is clearly seen in [O\,{\sc iii}]$\lambda$5007, extending to similar distances from the nucleus. For the other lines, the emission along the west spiral arm reaches approximately half the distance from the nucleus reached by the east spiral arm. As pointed out in Paper I, the VLA radio structure seems to correlate with the east spiral arm, here observed not only in H$\alpha$ but also in [O\,{\sc iii}] and [S\,{\sc ii}] line emission.

\subsubsection{Gas Excitation}
\label{gasex}

The [N\,{\sc ii}]6584/H$\alpha$ line-ratio values (top middle panel of Fig. \ref{razao}) are typical of AGN excitation, being characteristic of either Seyfert or LINER nuclear activity. The [O\,{\sc iii}]5007/H$\beta$ ratio map (bottom left panel of Fig. \ref{razao}) with values ranging from $\approx\,1.6\,$ to $\approx\,4.5\,$ is more characteristic of LINER activity, although there is varying excitation, with the nucleus, north and south regions as well as the farthest part of the west spiral arm showing the highest excitation, while the regions corresponding to the center of the east and beginning of the west spiral arms show the lowest excitation. 

\begin{figure}
\centering
\includegraphics[width=0.3\textwidth]{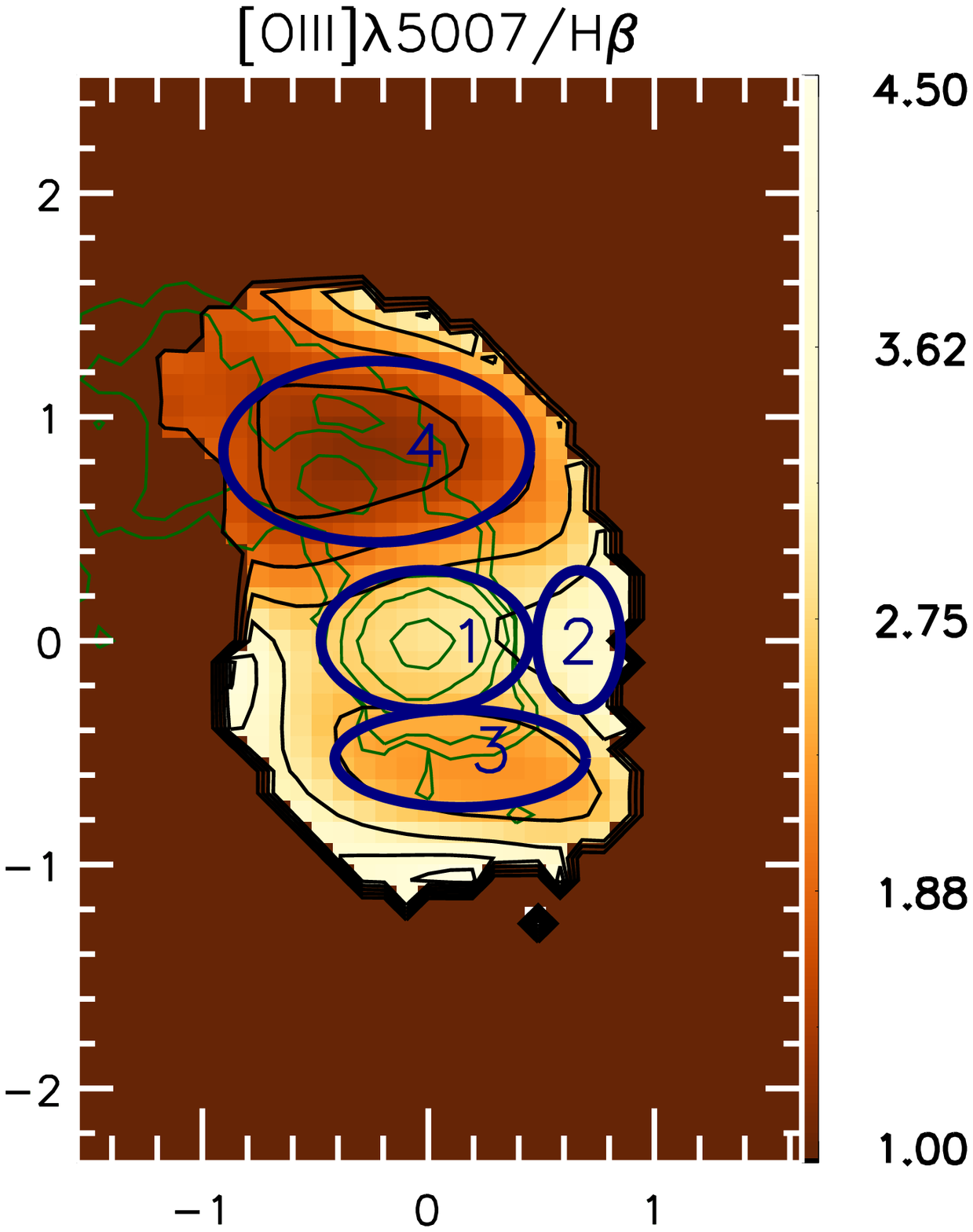}
\caption{[O\,{\sc iii}]5007/H$\beta$ ratio map displaying the regions represented by the symbols shown in Fig. \ref{diag}.}
\label{ratio}
\end{figure}

\begin{figure*}
\includegraphics[width=0.95\textwidth]{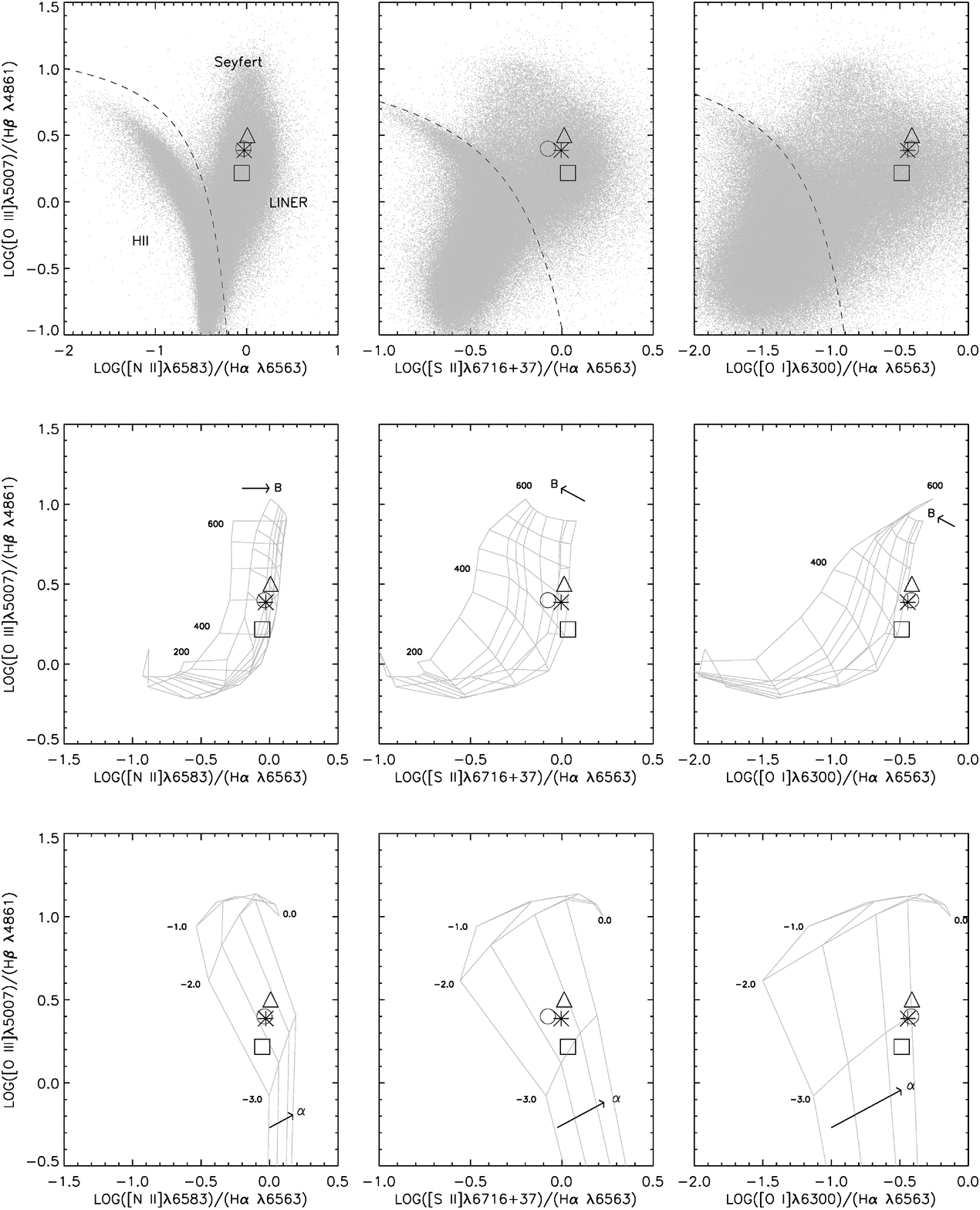}
\caption{Diagnostic diagrams (upper panels) in which the small gray dots represent ratios obtained from SDSS emission-line galaxies spectra. The dashed line divides the loci of HII-type galaxies (left) from that of AGNs (right).  Symbols represent regions displayed in the [O\,{\sc iii}]5007/H$\beta$ ratio map (Fig. \ref{ratio}). Region 1 is represented by an asterisk, region 2 by a triangle, region 3 by a circle and region 4 by a square. Shock + precursor line ratio grids from \citep{allen08} are displayed in the central panels (gray lines) in comparison to Arp 102B ratio values. The grids are displayed with shock velocities covering the range of 200 to 600 \kms in steps of 50 \kms and magnetic fields of 10$^{-4}$, 0.5, 1.0, 2.0, 3.23, 4.0, 5.0 and 10.0 $\mu$G. Dusty photoionization line ratio grids from \citep{groves04I,groves04II} are displayed in the bottom panels (gray lines) also in comparison to Arp 102B ratio values. The grids are displayed with power-law slope $\alpha$ values of $-2.0$, $-1.7$, $-1.
4$ and $-1.2$ (the black arrow shows the direction of increasing values) and ionization parameter $U$ values, in logarithmic scale, of $0.0$, $-1.0$, $-2.0$, $-3.0$ and $-4.0$ (the latter being actually below the figure limits).}
\label{diag}
\end{figure*}

In order to better map the gas excitation, we have built the diagnostic diagrams \citep{baldwin81} [O\,{\sc iii}]/H$\beta$ vs. [N\,{\sc ii}]/H$\alpha$, [O\,{\sc iii}]/H$\beta$ vs. [S\,{\sc ii}]/H$\alpha$ and [O\,{\sc iii}]/H$\beta$ vs. [O\,{\sc i}]/H$\alpha$, shown in the upper panels of Fig. \ref{diag}. Gray dots represent emission-line ratios for thousands of galaxy spectra in the {\it Sloan Digital Sky Survey} (SDSS) data \citep{cid09}. Large symbols represent the Arp102B data, shown as average line ratios within the regions identified in Fig. \ref{ratio}: region 1 -- corresponding to an oval of 340\,pc\,$\times$\,490\,pc around the nucleus -- is represented by an asterisk; region 2 -- oval of 200\,pc\,$\times$\,290\,pc at 290\,pc north-west of the nucleus -- is represented by a triangle; region 3 -- an oval of 590\,pc\,$\times$\,200\,pc at 250\,pc west of the nucleus, at the beginning of the west spiral arm --  is represented by a circle and region 4 -- oval of 690\,pc\,$\times$\,440\,pc at 440\,pc east of the nucleus, in the middle of the east spiral arm -- is represented by a square. Symbol sizes are of the order of the error bars, estimated by the standard deviation of the ratios in the regions. Table \ref{regions} shows the average integrated flux and average velocity dispersion values for each emission-line for each region plotted  in the diagnostic diagrams. The errors were derived from the Monte Carlo simulations.

\begin{table}
   \centering   
   \caption{\it Average integrated flux and velocity dispersions values for each region displayed in the diagnostic diagrams.}
   \begin{tabular}{|c|c|c|c|c|c|c|c|c|c|} 
      \hline \hline
      Emisson-line & Integrated flux & Velocity dispersion \\
      & (10$^{-15}$ erg cm$^{-2}$ s$^{-1}$) & (\kms) \\
      \hline
      &  Region 1 & \\
      \hline
      H$\beta$ & 0.86 $\pm$ 0.06 & 201.07 $\pm$ 6.09 \\
      $[$O\,{\sc iii}$]\lambda$5007 & 2.21 $\pm$ 0.10 & 193.66 $\pm$ 4.18 \\
      $[$O\,{\sc i}$]\lambda$6300 & 1.35 $\pm$ 0.07 & 195.21 $\pm$ 3.73 \\
      H$\alpha$ & 3.49 $\pm$ 0.25 & 196.30 $\pm$ 4.55 \\
      $[$N\,{\sc ii}$]\lambda$6584 & 3.22 $\pm$ 0.20 & 195.56 $\pm$ 4.19 \\
      $[$S\,{\sc ii}$]\lambda$6717 & 1.44 $\pm$ 0.05 & 171.99 $\pm$ 1.39 \\
      $[$S\,{\sc ii}$]\lambda$6731 & 1.47 $\pm$ 0.05 & 171.99 $\pm$ 1.39 \\
      \hline
      &  Region 2 & \\
      \hline
      H$\beta$ & 0.21 $\pm$ 0.03 & 152.78 $\pm$ 10.66 \\
      $[$O\,{\sc iii}$]\lambda$5007 & 0.50 $\pm$ 0.05 & 189.02 $\pm$ 9.55 \\
      $[$O\,{\sc i}$]\lambda$6300 & 0.27 $\pm$ 0.02 & 184.56 $\pm$ 7.85 \\
      H$\alpha$ & 0.68 $\pm$ 0.04 & 166.08 $\pm$ 4.48 \\
      $[$N\,{\sc ii}$]\lambda$6584 & 0.67 $\pm$ 0.04 & 190.60 $\pm$ 4.89 \\
      $[$S\,{\sc ii}$]\lambda$6717 & 0.32 $\pm$ 0.02 & 167.12 $\pm$ 3.13 \\
      $[$S\,{\sc ii}$]\lambda$6731 & 0.31 $\pm$ 0.02 & 167.12 $\pm$ 3.13 \\
      \hline
      &  Region 3 & \\
      \hline
      H$\beta$ & 0.23 $\pm$ 0.03 & 145.42 $\pm$ 9.11 \\
      $[$O\,{\sc iii}$]\lambda$5007 & 0.55 $\pm$ 0.05 & 141.33 $\pm$ 6.30 \\
      $[$O\,{\sc i}$]\lambda$6300 & 0.33 $\pm$ 0.02 & 170.73 $\pm$ 5.91 \\
      H$\alpha$ & 0.90 $\pm$ 0.05 & 157.35 $\pm$ 3.68 \\
      $[$N\,{\sc ii}$]\lambda$6584 & 0.82 $\pm$ 0.04 & 165.73 $\pm$ 3.70 \\
      $[$S\,{\sc ii}$]\lambda$6717 & 0.41 $\pm$ 0.02 & 163.36 $\pm$ 2.64 \\
      $[$S\,{\sc ii}$]\lambda$6731 & 0.39 $\pm$ 0.02 & 163.36 $\pm$ 2.64 \\
      \hline
      &  Region 4 & \\
      \hline
      H$\beta$ & 0.27 $\pm$ 0.03 & 159.09 $\pm$ 9.76 \\
      $[$O\,{\sc iii}$]\lambda$5007 & 0.49 $\pm$ 0.04 & 150.08 $\pm$ 7.49 \\
      $[$O\,{\sc i}$]\lambda$6300 & 0.31 $\pm$ 0.02 & 172.17 $\pm$ 7.15 \\
      H$\alpha$ & 0.87 $\pm$ 0.05 & 164.78 $\pm$ 4.20 \\
      $[$N\,{\sc ii}$]\lambda$6584 & 0.79 $\pm$ 0.04 & 165.99 $\pm$ 4.23 \\
      $[$S\,{\sc ii}$]\lambda$6717 & 0.45 $\pm$ 0.02 & 162.74 $\pm$ 2.76 \\
      $[$S\,{\sc ii}$]\lambda$6731 & 0.43 $\pm$ 0.02 & 162.74 $\pm$ 2.76 \\
      \hline
   \end{tabular}
   \label{regions}
\end{table}

The dashed lines, obtained from \citet{kewley06}, are the dividing lines between HII-type galaxies (left) and active galaxies (right). Higher [O\,{\sc iii}]/H$\beta$ values in the AGN side characterize Seyfert nuclei, and lower values characterize LINERs. Arp\,102B ratios are clearly located in the AGN side in the three diagnostic diagrams. The [O\,{\sc iii}]/H$\beta$ ratio values are mostly in the LINER region, although very close to the Seyfert loci, in particular for the extranuclear region 2, in agreement with the LINER/Seyfert 1 classification. The nuclear (region 1) line ratios are very similar to those of region 3 (beginning of the west arm), and are closer to the LINER region, while region 4, in the middle of the east arm is in the LINER region. 

Although the lower values of the [O\,{\sc iii}]/H$\beta$ ratio map in the east spiral arm (and in the beginning of the west spiral arm) could be due to the presence of HII regions, the other line ratios imply that the loci of the corresponding data points in the diagrams do not support this interpretation, favoring instead ionization by a diluted continuum and/or the contribution of shocks along the east and west spiral arms.

In an attempt to distinguish the contribution of these two excitation mechanisms, we compare our line ratios with those predicted by shock and photoionization models from the literature. In a shock-ionization scenario, if the expanding shock wave velocity is high enough ($\approx 200$ \kms), the recombination region behind the shock front gives origin to photons that travel upstream to ionize also a region ahead of the shock, which is called the precursor region. In the central panels of Fig. \ref{diag} we present shock + precursor line ratio grids (gray lines) from the MAPPINGS III shock models \citep{allen08}. The grids are displayed with different values of shock velocity and magnetic field. Even though there is a degeneracy in the values of the [O\,{\sc iii}]/H$\beta$ vs. [N\,{\sc ii}]/H$\alpha$, Arp 102B data can be reproduced by these models for shock velocities of 400 to 500 \kms and low magnetic fields of 10$^{-4}$ to 2.0.

The bottom panels of Fig. \ref{diag} show line ratio grids (gray lines) from the MAPPINGS III photoionization models \citep{groves04I,groves04II} for dusty emission-line regions ionized by a power-law continuum. The grids are displayed with different values of power-law slope $\alpha$ and ionization parameter $U$ values. Other parameters of the models were the gas density -- fixed at $n_H = 1000$\,cm$^{-3}$ (approximate value implied by our [S\,II] ratio map) and gas metallicity -- fixed at twice the solar value. The data in [O\,{\sc iii}]/H$\beta$ vs. [N\,{\sc ii}]/H$\alpha$ and [O\,{\sc iii}]/H$\beta$ vs. [S\,{\sc ii}]/H$\alpha$ diagrams are well reproduced by models with $-1.7\le\alpha\le-1.4$ and $-3.0 < log U < -2.5$. However, the data in the [O\,{\sc iii}]/H$\beta$ vs. [O\,{\sc i}]/H$\alpha$ are more consistent with a flatter power-law slope and a slightly lower ionization parameter, $log U \approx -3$, or in other words, the observed [O\,{\sc i}]/H$\alpha$ ratio is stronger than predicted by the 
models that account for the other line ratios. A possible explanation is that there is a contribution from gas that is photoionized by a partially absorbed continuum spectrum. Absorption of photons with energies near the Lyman limit will effectively produce a harder continuum, which will tend to result in a larger semi-ionized zone, where [O\,{\sc i}] is produced.

\begin{figure*}
\includegraphics[width=\textwidth]{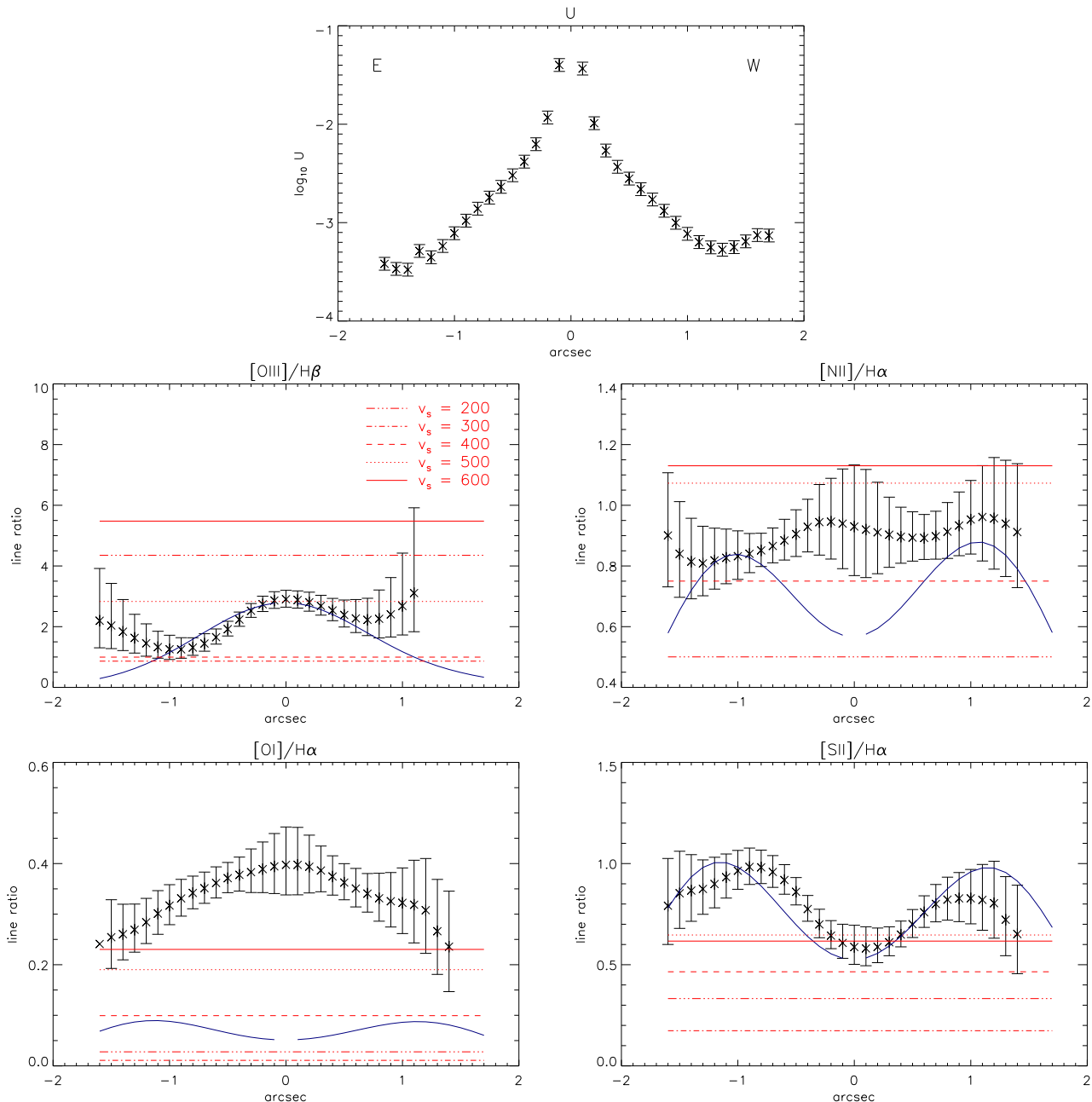}
\caption{Top panel: Photoionization parameter $U$ calculated along a pseudo-slit at PA$\approx 90^{\circ}$ as a function of distance from the nucleus. The middle and bottom panels show the observed emission-line ratios in black crosses, along with shock + precursor model values from \citet{allen08} (red lines) and photoionization model values from \citet{groves04I,groves04II} (blue lines) for the values of $U$ shown in the upper panel. The power-law slope $\alpha$ assumed was $-2$, the gas density was fixed $n_H = 1000$\,cm$^{-3}$, and the gas metallicity Z was assumed to be 2Z$_\odot$. A Gaussian smoothing (representing the PSF) was applied to the spatial variation of the photoionization model values.}
\label{photo}
\end{figure*}

In order to further investigate if photoionization only can explain our measurements, we have calculated the ionization parameter as a function of distance from the nucleus assuming that the emitting gas is photoionized solely by a central source with an emission rate of ionizing photons $Q(H^0)$. We calculate the ionization parameter $U$ at a distance $r$ from the source as \citep{ostfer06}:

\begin{equation}
U = {\frac{Q(H^0)}{4 \pi r^2 c n_H}} {\frac{1}{C}} \, ,
\label{}
\end{equation}

\noindent
where $n_H$ is the gas density and $C$ is the covering factor. In order to obtain $U$ as a function of distance from the nucleus, we have considered a ``pseudo-slit'' along position angle $\approx 90^{\circ}$ (along the east and west spiral arms). $Q(H^0)$ can be estimated through the total luminosity measured in the H$\beta$ emission-line (see eq. 5.34 of \citet{ostfer06}) and $n_H$ is calculated using the sulfur lines ratio at the corresponding radial distance along the pseudo-slit (see Sec. \ref{densi}). The upper panel of Fig. \ref{photo} displays the photoionization parameter $U$ as a function of the projected distance from the nucleus, for a covering factor of $0.2$ (see discussion below). Error bars are due to propagated uncertainties in flux measurements of the [S\,{\sc ii}] and H$\beta$ emission-lines. $U$ has Seyfert-like values close to the nucleus ($log U \approx -1.2$), which decrease to LINER-like values ($log U \approx -3.5$) beyond $1\farcs$ from the nucleus. The central and lower panels of 
Fig. \ref{photo} show the observed line ratios along the pseudo-slit in black crosses, with error bars also due to flux measurement  uncertainties. In red lines we display the expected value of each line ratio from the shock models from \citet{allen08}, with several shock velocities, identified by the labels in the middle left panel. In blue we show the line ratios we have obtained using the models of \citet{groves04I,groves04II} for dusty emission-line regions for the ionization parameter values we have calculated (upper panel). In order to reproduce the model values, we verified by trial and error that the parameters which best reproduce the spatial variation of most line ratios are: $n_H = 1000$\,cm$^{-3}$, $\alpha = -2$ and $C = 0.2$. In addition, the spatial distribution of the modeled line ratios was convolved with a Gaussian profile representing the Point-Spread Function.

The line ratio values observed for [O\,{\sc iii}]/H$\beta$ and [S\,{\sc ii}]/H$\alpha$ seem to follow the same pattern as that presented by the smoothed spatial variation of the photoionization model values. In the case of [N\,{\sc ii}]/H$\alpha$, the pattern is quite different between the observed and modeled line ratios, indicating a contribution of another excitation mechanism, probably shocks, to reproduce the nuclear line ratios. A contribution from shocks could also explain the drastic difference seen for [O\,{\sc i}]/H$\alpha$, which also displays a different pattern and much higher values from those of the photoionization models. This is indeed the most shock-dependent ratio.

The behavior of the line ratios as a function of $r$ thus suggest the contribution of both shocks and photoionization to the gas excitation. This is consistent with the fact that this galaxy is classified as a LINER/Seyfert 1. The shock velocities implied by the models range between $400$ and $600$ \kms, while the power-law index suggested by the photoionization models ranges between $-2.0$ and $-1.4$. The measured centroid velocities and velocity dispersions ($\le 100$ and  $\le 200$ \kms, respectively; see Fig. \ref{velocidade}) are much smaller than the shock velocities implied by the models. This argues against fast shocks and hence favours AGN photoionization as the dominant ionization mechanism. However, we cannot rule out the possibility of high velocity shocks associated with bulk flows in or near the plane of the sky. Indeed, we argue that the velocity field shows evidence of just such an outflow associated with the radio jet along the east spiral arm (Sections \ref{gaskin} and \ref{outflow}).
 
\subsubsection{Reddening}
\label{redden}

We have used the observed H$\alpha$/H$\beta$ line ratio (Fig. \ref{razao}) to calculate the gas reddening. Adopting the \citet{cardelli89} reddening law, and assuming case B recombination \citep{ostfer06} we obtain

\begin{equation}
E(B-V) = 2.22\, \log\, \frac{(\frac{H\alpha}{H\beta})}{3.1} \, .
\label{aver}
\end{equation}

\begin{figure*}
\begin{center}
\includegraphics[width=\textwidth]{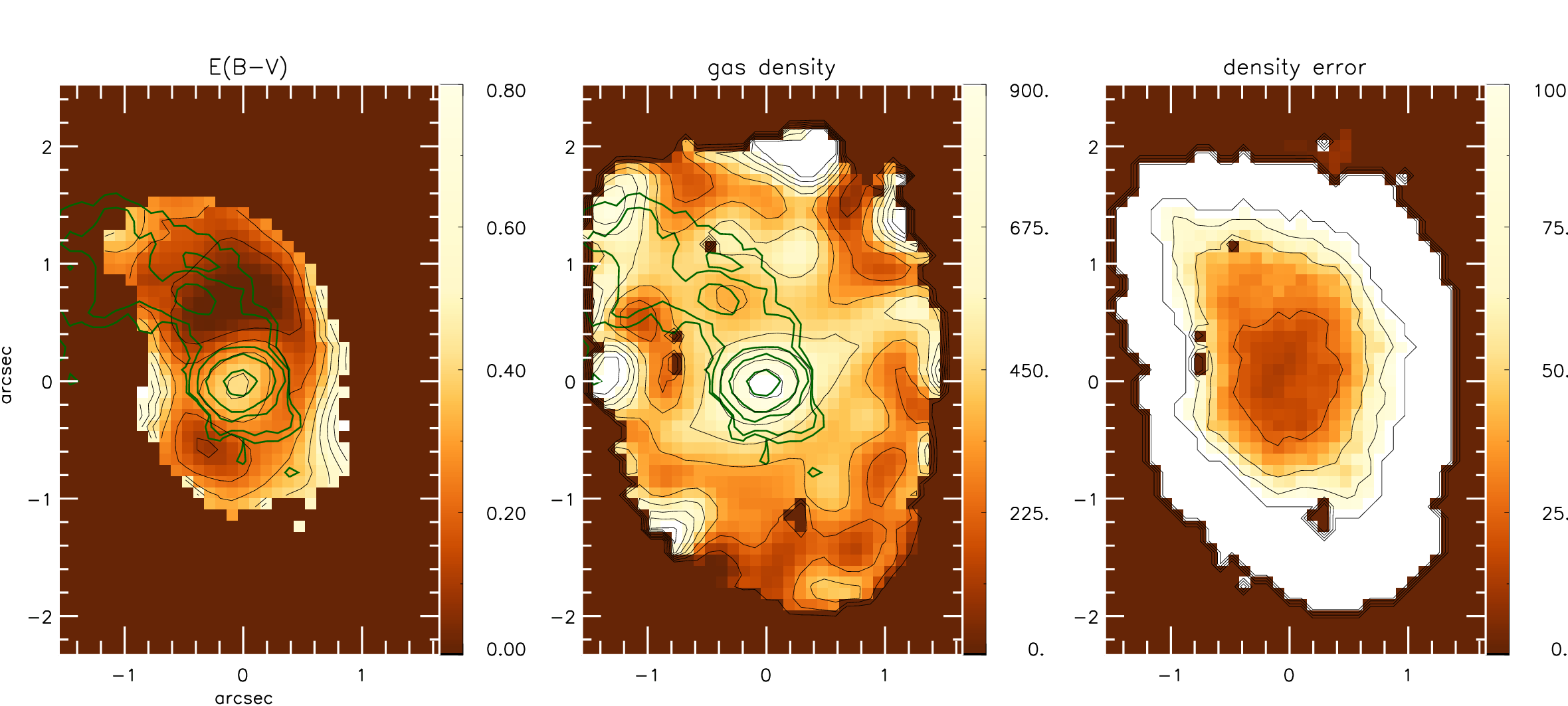}
\caption{Left panel: reddening map, with values of E(B-V) obtained from the H$\alpha$/H$\beta$ ratio. Middle panel: gas density map, derived from the [S\,{\sc ii}] ratio. Density units are cm$^{-3}$. Right panel: mean uncertainty values for the gas density, in percent scale.}
\label{red}
\end{center}
\end{figure*}

The corresponding reddening map is shown in the left panel of Fig. \ref{red}. The highest reddening values (E(B-V)$\geq$\,0.4) are observed at the nucleus and both to the south-east and north-west of it. The smallest reddening values, close to 0, are observed at $\approx$\,1\arcsec\, from the nucleus, in the middle of the east spiral arm, surrounded by regions with (E(B-V))$\approx\,0.25$, a value observed also in the beginning of the west spiral arm. This map suggests that the spirals -- and the radio jet -- may have cleared a channel in the beginning of the arms pushing away the dust along the way.

\subsubsection{Gas density}
\label{densi}

We have used the ratio of the [S\,{\sc ii}] emission lines (top right panel of Fig. \ref{razao}) to calculate the gas density \citep{ostfer06}, whose map is displayed in the middle panel of Fig. \ref{red}. The temperature used to estimate the gas density is $10,000$\,K. The average gas density over all IFU FOV is $n_e = 440^{+290}_{-160}$\,cm$^{-3}$. The highest density values of $\approx$\,900\,cm$^{-3}$ are observed at the nucleus, approximately co-spatial with the location of highest reddening. Away from the nucleus, the highest densities are observed along the east spiral arm, with an average value of about 500\,cm$^{-3}$, but presenting two knots of higher density values at its northern and eastern borders. These knots seem to bound the region covered by the bent radio jet. In particular, a radio knot at 0$\farcs$8 east of the nucleus occurs adjacent (below in Fig. \ref{red}) to a high-density knot, suggesting that this high density knot may have deflected the radio jet, leading to its bent appearance. 
Error values for the gas density are displayed in the right panel of Fig. \ref{red}. The density errors, in percent scale, are highest towards the borders of the IFU-FOV. These uncertainties originate from flux errors measured with the Monte Carlo iterations, illustrating that measurements close to the borders of the FOV are not reliable enough to estimate the gas density. However, the errors are small enough around the nucleus and in the high density knot, that they do not compromise our interpretation in these regions.

The presence of high density knots along the east spiral arm, as well as the radio knot 0$\farcs$8 east of the nucleus, combined with the high velocity dispersions observed along the east arm up to the radio knot (right top and middle panels of Fig. \ref{velocidade}) support at least some contribution from shocks to the gas ionization in this region, leading to LINER-like emission-line ratios.

The gas excitaton with LINER-like line ratios could be due to shocks or diluted radiation. Although the velocity dispersion maps shows no unusual high values in the region, the scenario including shocks cannot be ruled out, once the dispersion is not negligible along the spiral arms. We have also considered the possibility that the low [O\,{\sc iii}]/H$\beta$ values were due to the presence of a star forming region in the east spiral arm, but this hypothesis can be discarded, as the high [N\,{\sc ii}]6584/H$\alpha$ and [O\,{\sc i}]6584/H$\alpha$ ratios are much larger than those observed in HII regions.

\subsection{Gas Kinematics}
\label{gaskin}

Fig. \ref{velocidade} shows that the centroid velocity fields suggest rotation, with redshifts to the west and blueshifts to the east. Nevertheless, the isovelocity curves are distorted relative to the well known ``spider diagram'' (characteristic of rotation), and a possible interpretation is a combination of gas rotation and outflow. The apparent S-shape of the central isovelocity curves could be due to the presence of what seems to be a ``blob'' of blueshifted line emission at $\approx 0\farcs7$ south-east of the nucleus (to the left of the nucleus in Fig. \ref{velocidade}) with negative velocities in excess of $-50$\,\kms in the H$\alpha$ centroid velocity map. This ``blob'' could be interpreted as due to an outflow which could also explain the high velocity dispersions observed at its location. Although not conspicuous in the channel maps as a detached structure, the blueshifted channel maps between velocities $\approx\,-150$ and $-50$\,\kms do show an extension to the south-east, at the location of 
this blob.

Another interesting feature of the centroid velocity maps is a region of lower blueshifts -- co-spatial with the radio jet -- surrounded by higher blueshifts. In the channel maps we see that although the jet region is dominated by blueshifts, there are also some redshifts observed at the same locations. This could be due to a broad outflow launched close to the plane of the sky, with the front part approaching and the back part receding, but slightly tilted toward us, so that the centroid velocities would still show blueshifts, but lower than the surrounding rotation velocities. Thus the highest blueshifts to the east and highest redshifts to the west could be due to rotation, while the intermediate blueshifts and redshifts to the east would include the contribution from the front and back walls of the outflow. 

\subsubsection{Velocity rotation modelling}

\begin{figure*}
\centering
\includegraphics[scale=0.35]{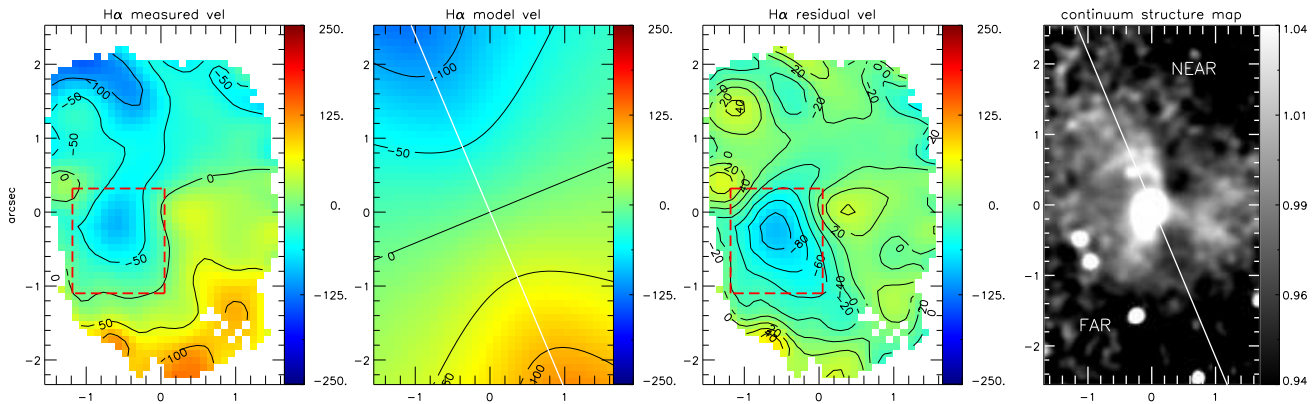}
\caption{Gas rotation model for H\,$\alpha$. Left: the measured centroid velocity map. Middle left: model of the rotation curve. Middle right: residual map. Right: structure map. The red dashed box illustrates the masked region due to high residual velocities, from a possible outflow. The white line displays the position of the line of nodes. Velocities are in \kms.}
\label{vmodf}
\end{figure*}

In order to try to separate the contributions of possible rotation and outflow to the velocity field, we fitted a rotation model to the H$\alpha$ centroid velocity map, assuming that the emitting gas is rotating in a central potential, as we have done in previous works \citep{barbosa06, riffel08, schnorr11}. In this kinematic model \citep{bertola91, kruit78} it is assumed that the gas has circular orbits in a plane, and the rotation field is given by

\begin{equation}
\begin{split}
v&_{mod} (R, \Psi) = v_{sys} + \\
&{\frac{A R\, cos (\Psi - \Psi_0)\, sin \theta\, cos^p \theta}{\{R^2[sin^2(\Psi - \Psi_0) + cos^2 \theta\, cos^2 (\Psi - \Psi_0)] + c_0^2\, cos^2 \theta\}^{p/2}}}
\end{split}
\label{vmode}
\end{equation}

\noindent
where $v_{sys}$ is the systemic velocity, $A$ is the centroid velocity amplitude, $\Psi_0$ is the major axis position angle, $c_0$ is a concentration parameter, $\theta$ is the angle between the disk plane and the sky plane, $p$ is a model fitting parameter (which is $\sim\,1$ for finite masses in a Plummer potential) and $R$ and $\Psi$ are the coordinates of each pixel in the plane of the sky. Fig. \ref{vmodf} shows in the first three panels the measured H$\alpha$ centroid velocity map, the resulting model of rotation curve and the residual between both, respectively. The red dashed box shown encloses the ``blueshifted blob'', which is found in the centroid velocity maps. Since this region seems to show the most significant deviations from a rotation pattern, we decided to mask it out of the fit. The resulting parameters are: $A \approx 268 \pm 8$ \kms, $v_{sys} \approx 7244$ \kms, $\Psi_0 \approx 88^{\circ}$ , $c_0 \approx 5$ arcsec and $\theta \approx 79^{\circ}$ (error bars in almost all parameters are 
smaller than unity, thus not shown here). We fixed the center of the rotation curve in $x_0 = 0$ and $y_0 = 0$ arcsec, with $p = 1$. The systemic velocity resulting from the modelling is close to the average value we adopted as systemic (Sec. \ref{ifukin}) and both are in agreement with \citet{devaucouleurs91} ($7245 \pm 34$\,\kms). Fig. \ref{vmodf} shows the observed velocity map in the left panel, the model velocity field in the center panel, and the residual between the two in the right panel.

The masked region is the only one with values larger than $40$\,\kms in the residual map. This indicates that our hypothesis of combined rotation and outflow in the blueshifted ``blob'' is a good representation of the data. A small redshifted region in the residual map (residuals of $\approx\,40$\,\kms), located to the right (north-west) of the nucleus, suggests the presence of a redshifted blob which could be the counterpart for the blueshifted ``blob'', in the hyphotesis they are due to a bipolar outflow.

Fig. \ref{vmodf} shows that the rotation model is a good representation of the gas centroid velocity kinematics, and that the disk seems to be very inclined. In order to try to determine which is the near and far sides of the disk, we have looked for possible signatures of dust obscuration, which is usually larger in the near side of an inclined disk.

Dust lanes can be highlighted via structure maps \citep{pogmar02, simoes07}, in which high contrast regions -- such as those resulting from dust  obscuration or high emission structures -- can be enhanced, appearing as dark and bright regions, respectively. The right panel of Fig. \ref{vmodf} shows the structure map we have obtained from the HST ACS continuum image shown in Fig. \ref{large}. A dark lane to the north-east of the nucleus running approximately east-west (passing to the right of the nucleus in the figure) could be due to extinction by dust in the near side of a highly inclined disk. 

The dust lane thus seems to support a high inclination for the rotating disk, as derived from our rotation model (79$^\circ$). And suggest that the spiral arms are not all in the disk, as they would not be visible at such high inclination. Our hypothesis is that the jet is launched close to the plane of the sky and of the line of nodes of the rotating disk, then suffers a deflection to high latitudes relative to the rotating gas disk, dragging along gas from the disk.

In order to have more elements to interpret the overall kinematics, we decided to apply a PCA analysis to the IFU data \citep{steiner09} in a similar way as described in the previous study by members of our group of the gas kinematics of the nuclear region of M81 \citep{schnorr11}.

\subsubsection{PCA Analysis}
\label{pcan}

The PCA has been applied directly to the calibrated datacube separating the information, originally presented in a system of correlated coordinates, into a system of uncorrelated coordinates ordered by principal components of decreasing variance. These new coordinates called ``eigenspectra''  reveal spatial correlations and anti-correlations in the emission-line data. The projection of the data onto these new coordinates produce images called ``tomograms'', which map the spatial distribution of the eigenspectra.

\begin{figure*}
\centering
\includegraphics[scale=0.37]{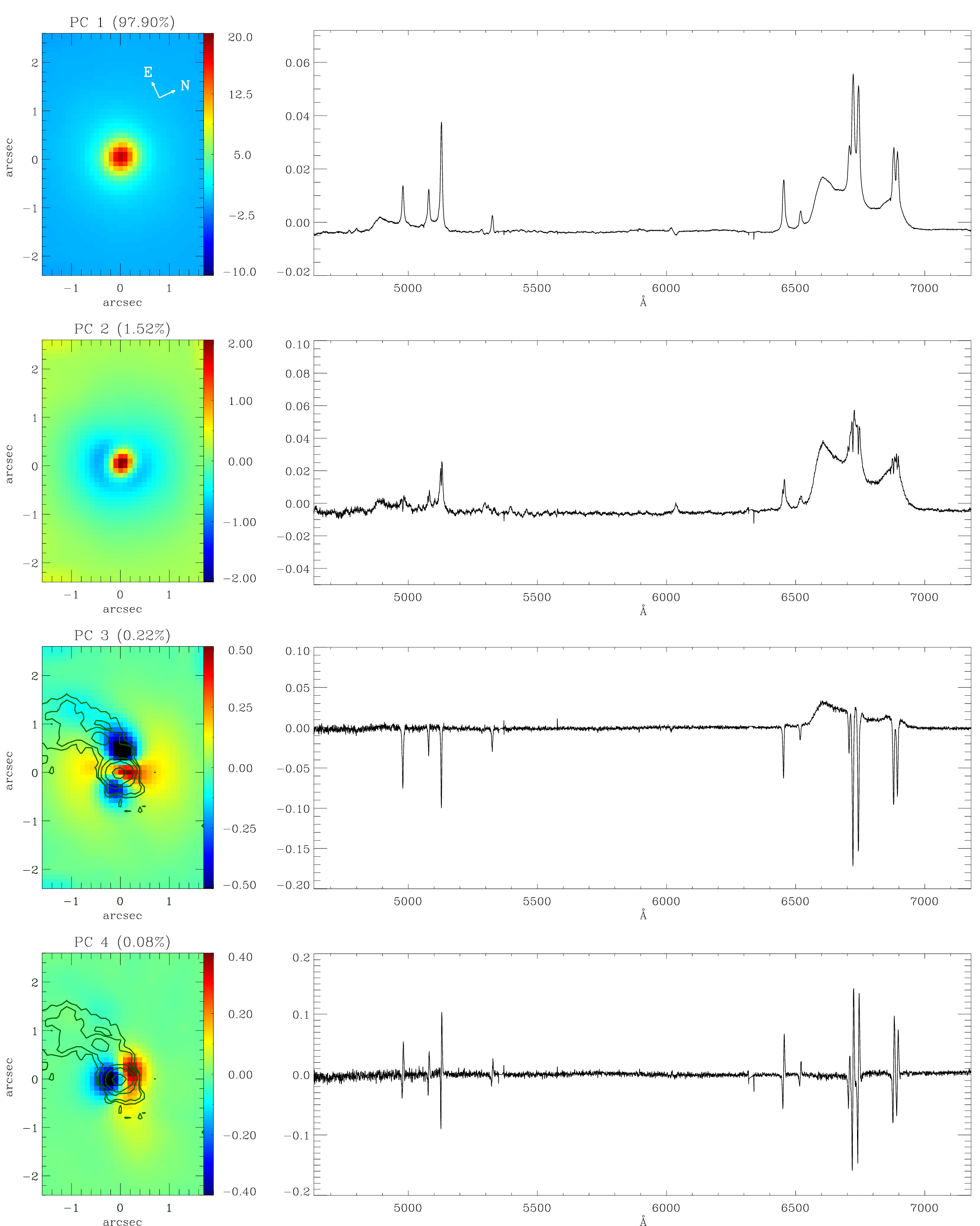}
\caption{PCA Analysis. From top to bottom: tomograms (left) and eigenspectra corresponding to PC1, PC2, PC3 and PC4, in decrescent order of variance.}
\label{pca}
\end{figure*}

The result of the PCA is shown in Fig. \ref{pca}. From top to bottom the panels show the tomograms (to the left) and respective eigenspectra (to the right), in order of decreasing variance. 

Component PC1, which carries the largest variance (97.9\%), shows an eigen-spectrum dominated by the AGN and the stellar continuum contribution, represented by the positive values in the eigenspectrum and showing a circularly symmetric spatial distribution in the corresponding  tomogram. 

Component PC2 is dominated by the contribution of the broad-line emission, originating in the nucleus -- the red region of the tomogram, being spatially anti-correlated with the stellar population, which originates from the blue region (circumnuclear region) in the tomogram. The anti-correlation is revealed by the inverted stellar population features (which appear in emission) in the eigenspectrum (e.g. the NaI$\lambda$5890 and Mg$\lambda$5175 absorption features characteristic of stellar populations). This eigenspectrum shows little contribution from the [N\,{\sc ii}], H$\alpha$ and [S\,{\sc ii}] emission lines, being remarkably similar to the STIS nuclear spectrum (see Fig. \ref{large}). We interpret this as being due to the smaller aperture of the STIS spectrum, which is dominated by the broad H$\alpha$ component. The PCA has thus separated this nuclear component from the more extended narrow line emission in our datacube, even though our IFU data has a much smaller angular resolution than the STIS data. 
This demonstrates the power of the PCA technique \citep{steiner09,ricci11}. This eigenspectrum also shows that there is some [O\,{\sc iii}] and [O\,{\sc i}] emission originating in the same region as that of the broad H$\alpha$, or at least much closer to the nucleus than the other emission lines.

The PC3 eigenspectrum shows narrow lines anti-correlated (in absorption) with the broad component of H$\alpha$ (in emission). This broad component originates in the red region of the tomogram, while the narrow lines correspond to blue regions in the tomogram. The narrow emission lines thus originate in two blobs, one more extended to the east and one less extended to the west of the nucleus, coinciding with the location of the beginning of the spiral arms. Notice that there is no velocity gradient in the emission lines, indicating that the two blobs originate in gas at similar velocities. One possibility is that these blobs are due to gas being ejected close to the plane of the sky. This result supports the hypothesis put forth in the previous section that the jet is launched close to the plane of the sky (approximately along the major axis of the rotating disk) and that these blobs correspond to the inner parts of the outflows pushed by the radio jet.

The PC4 eigenspectrum shows the narrow lines with components in blueshift (in absorption), spatially anticorrelated with components in redshift (in emission). The tomogram shows that the components in blueshift originate to the south-east of the nucleus, while the redshifted component originates to the north-west of the nucleus. The blueshifted region is very well aligned with the blueshifted ``blob''. The counterpart structure observed in redshift has also been previously seen in the residual map of the velocity rotation model. Both the blueshifted and redshifted regions show high velocity dispersions, and we propose that they are due to a bipolar outflow, almost perpendicular to the radio jet. Another possible interpretation for such a feature is a compact rotating disk, smaller than the one observed in the centroid velocity maps. The disk would be almost perpendicular to the radio jet, suggesting that it could be a mechanism of fuelling the AGN.

The broad component that appears in the PC3 eigenspectrum is extended in an apparent bipolar distribution which is centered on the nucleus and oriented approximately SE-NW, with the NW side the stronger. The most plausible explanation for this feature is that it represents scattered light from the broad-line region. It is also notable that this PC3 feature is approximately co-spatial, though slightly misaligned, with the structure in PC4, which we identify as a bipolar outflow. This suggests that the broad PC3 feature is associated with scattering of broad-line emission in the outflow, with the spatial distribution indicating that the strongest scattering occurs in the redshifted (receding) side of the flow. The misalignment between the PC3 and PC4 structures also suggests that in both the blue and redshifted sides, the scattering is concentrated along one edge of the flow (the eastern edge of the blueshifted flow and the western edge of the redshifted flow, respectively). This may be due to a density gradient across the flow. The outflow axis is approximately perpendicular to the line of nodes of the large scale rotating gas disk, the northern side of which we believe to be the nearside. This implies that the blue and redshifted sides of the outflow are in front of and behind the disk, respectively, with their the eastern and western edges being closest to the disk. The interpretation of the broad-line scattering obviously argues against the scenario of a rotating disk for PC3.

Support for the bipolar outflow interpretation comes from spectropolarimetric observations of Arp\,102B \citep{antonucci96,corbett98}, which show that the broad H$\alpha$ line is polarized with a position angle ($\approx 103^\circ$) closely aligned with that of the radio jet. As scattering produces polarization perpendicular to the scattering plane (i.e., the plane containing both the incident and scattered rays), this is consistent with scattering in a structure elongated perpendicular to the radio jet axis, as suggested by the PC3 eigenspectrum. \citet{chen89_2} modelled the asymmetric double-peaked broad H$\alpha$ line profile as emission from a relativistically rotating disk. Adopting this model to represent the intrinsic line profile, \citet{corbett98} showed that the shape of the broad H$\alpha$ profile in polarized flux is well reproduced by scattering in a bipolar outflow (see their Fig. 4). \citet{corbett98} assumed the scattering outflow to be aligned with the radio axis, and noted that in this case the polarization position angle should be perpendicular to the jet axis, contrary to what is observed. However, the observed alignment can be understood if the bipolar outflow is, in fact, perpendicular to the jet axis.

We conclude that not only the PC3 and PC4 eigenspectra but also the broad H$\alpha$ polarization are strongly suggesting a bipolar outflow which emits narrow lines and scatters nuclear broad-line emission. Remarkably, the axis of this outflow, as determined both from the PCA tomograms and from the polarization angle, must be approximately perpendicular to that of the radio source.

We note that \citet{ricci11} invoked a somewhat similar structure to explain the PCA eigenspectra of their IFU data cube covering the center of the elliptical galaxy NGC7097. In that case, a PCA eigenspectrum identified with the LINER nucleus is attributed to scattering in an ionization bicone aligned with the axis of a rotating disk. Unlike Arp\,102B, however, there is no (known) radio jet and no optical polarimetry from which additional geometrical constraints might be obtained.

Two sets of perpendicular outflows have been observed previously in other targets, such as Markarian 6 \citep{kharb06}. This could be the case here: we are seeing a new outflow perpendicular to the first due to precession of a nuclear jet. In fact, Seyfert galaxies have short-lived outflows ($10^4$-$10^5$ years) since the SMBH spin-axes change frequently and the accretion disks are short-lived \citep{sikora07}.

\subsection{One-sidedness mechanisms in radio jets}

In Paper I, we attributed the one-sidedness of the radio jet of Arp 102B to relativistic boosting. Even though our IFU observations favor a scenario where the jet is being launched close to the plane of the sky, we should still consider the interpretation of relativistic boosting, since we observe a bending of the jet about $\approx\,0\farcs5$ from the nucleus, probably towards our line-of-sight.

It is possible, if the twin jets have different intrinsic powers, that an asymmetric density distribution in the surrounded interstellar medium can cause a disruption of the counter-jet. This could be supported by our density map (Fig. \ref{red}), as long as this occurs within few hundred parsecs from the nucleus, where the gas density is highest. Nevertheless, our data seems unsufficient to distinguish between these models. Further radio observations, perhaps in several frequencies, would better constrain these possibilities.

\subsection{Mass of the emitting gas}

We estimated the mass of the ionized gas as:

\begin{equation}
M = V f n_e m_p\, ,
\label{eq2}
\end{equation}

\noindent
where $V$ is the volume of the emitting region, $f$ is the filling factor, $n_e$ is the electron density and $m_p$ is the proton mass.

From \citet{ostfer06}, assuming case B recombination, we obtain:

\begin{equation}
V\, f = 8.1 \times 10^{59}\,{\frac{L_{41}(H\beta)}{n_3^2}}\,\mathrm{cm^{-3}}\, ,
\label{eq1}
\end{equation}

\noindent
where $L_{41}(H\beta)$ is the H$\beta$ luminosity in units of $10^{41}$ erg s$^{-1}$ and $n_3$ is the electron density in units of $10^3$ cm$^{-3}$. The estimated mass of the emitting region comes from the introduction of equation \ref{eq1} into equation \ref{eq2}, which gives:

\begin{equation}
M \approx 7 \times 10^5\,{\frac{L_{41}(H\beta)}{n_3^2}} M_{\odot}\,.
\end{equation}

\noindent
This calculation can be seen in more detail in \citet{peterson97}.

In order to calculate the total H$\beta$ luminosity, we needed to correct the integrated H$\beta$ flux, $F(H\beta)$, for reddening. For the assumed distance of 104.9 Mpc and using the reddening law of \citet{cardelli89}, we have:

\begin{equation}
\begin{split}
L(H\beta) = 4\pi d^2 F(H\beta)10^{C(H\beta)} \\
= 1.4 \pm 1 \times 10^{40} \, &\mathrm{erg\,s^{-1}}\, ,
\end{split}
\end{equation}

\noindent
where $C(H\beta)$ is the estimated interstellar extinction coefficient. 

Using an average electron density $n_3 = 0.44^{+0.29}_{-0.16}$\,cm$^{-3}$ we obtain a mass of the emitting gas of $M = 3.1 \pm 2.7 \times 10^5 M_\odot$.

\subsection{Mass outflow rate}
\label{outflow}

In the analysis of the gas kinematics, we found evidence for an outflow associated with the radio jet along the east spiral arm and reaching $\approx 700$\, pc ($1\farcs5$) from the nucleus, and another one causing the blueshifted ``blob'' at $\approx 350$\,pc ($0\farcs6$) south-east from the nucleus. In this section we estimate the mass outflow rate for the first, stronger outflow, as it seems to have a much larger associated mass.

In order to estimate the mass outflow rate, we need to adopt geometric parameters for the outflow. The derivation of such parameters is a hard task, once there are projection effects and the outflow geometry is not clear. We have assumed a truncated conical geometry (conical frustrum) in which the nucleus is at the smaller base, which is a circle with a diameter of $0\farcs6$. The larger base (at the outer end of the outflow) is a circle with a diameter of $1\arcsec$, distant $1\farcs5$ from the smaller base. The cone from which the frustrum originates has an aperture of $15^\circ$. The area of the top base, crossed by the outflow is $A = \pi r^2$, where $r = 1\farcs0$ ($= 490$ pc), thus $A = 1.8 \times 10^{42}$\,cm$^{-2}$. The mass outflow rate can be calculated as \citep{rifsto11}:

\begin{equation}
\dot{M}_{out} = m_p\,n_e\,v_{out}\,A\,f\, ,
\end{equation}

\noindent
where, $m_p = 1.7 \times 10^{-24}$\,g is the proton mass, $n_e = 440^{+290}_{-160}$\,cm$^{-3}$ is the electron density (calculated as explained in the last section), $v_{out}$ is the velocity of the outflow perpendicular to $A$ at $\approx 1\farcs5$ east of the nucleus and $f$ is the filling factor.

The observed geometry and low velocity for the outflow, as well as the PCA results, suggest that the angle between the outflow and the plane of the sky is small. The fact that both blueshifts and redshifts are observed in the region of the outflow (see Fig. \ref{cmHalfa}), with a small excess of blueshifts suggests that the outflow is slightly tilted towards us, but part of the outflow must be moving away from us. We estimate the velocity and orientation  of the outflow using Fig. \ref{cmHalfa}, considering that the highest  blueshifts ($\approx -160$\,\kms) to the east originate approximately in the front part of the outflow while the highest redshifts ($\approx 130$\,\kms) originate in the back part of the outflow. We have considered the truncated conical geometry described above and have assumed that the velocity is the same all over the outflow. The highest blueshifts and redshifts are considered to be the projections (along the line-of-sight) of the velocity along the front and back parts of the outflow,
 respectively. Under these assumptions, we obtain an orientation of the outflow of only $\approx$\,$0.5^\circ$ relative to the plane of the sky towards us, and a velocity of $v_{out} = 1100$\,\kms for the outflow. The total flux in the region of the outflow is $F(H\beta) \approx 5.0 \times 10^{-15}$\,erg\,cm$^{-2}$\,s$^{-1}$ with a luminosity of $L(H\beta) = 4\pi d^2\,F(H\beta) \approx 6.6 \times 10^{39}$\,erg\,s$^{-1}$. With the geometry assumed, which results in $V \approx 2.7 \times 10^{63}$\,cm$^3$, we obtain a filling factor of $f \approx 1.8 \times 10^{-4}$, using equation \ref{eq1}. We then obtain a mass outflow rate of $\dot{M}_{out} \approx 0.32 \, \mathrm{M_\odot\, yr^{-1}}$.

In order to estimate an uncertainty for the mass outflow rate, we have used as alternative geometry for the outflow, namely a conical one, considering that the ``broad base'' at the nucleus may be due to smearing by the seeing. Assuming a cone with $1\arcsec$ base diameter and $1\farcs5$ of height, the aperture angle is $\approx 37^\circ$. Repeating the calculations described above, we obtain an angle between the cone axis and the plane of the sky of $1.6^\circ$, and an outflow velocity of $v_{out} = 470$\,\kms. The volume of the assumed geometry is $V \approx 1.4 \times 10^{63}$\,cm$^3$. This values result in a filling factor of $f \approx 3.6 \times 10^{-5}$, about twice the one calculated using a truncated cone as the outflow geometry. Finally we obtain a mass outflow rate of $\dot{M}_{out} \approx 0.26 \, \mathrm{M_\odot\, yr^{-1}}$.

An alternative interpretation for the blueshifts and redshifts observed along the east arm is that they are solely due to lateral expansion of ambient gas in a cocoon around the radio jet \citep{holt08, su12}. Blueshifts and redshifts would originate from the walls of the cocoon moving towards and away from the observer, in agreement with those observed in the channel maps. Also, in this case, the velocity of the outflow should be smaller than previously estimated. Under the assumption that the speed of the forward expansion along the radio jet is approximately the same as that of the lateral expansion we have calculated the mass outflow rate for $v_{out} \approx 150$\,\kms of $\dot{M}_{out} \approx 0.04 \, \mathrm{M_\odot\, yr^{-1}}$ assuming the geometry to be a truncated cone, and $\dot{M}_{out} \approx 0.08 \, \mathrm{M_\odot\, yr^{-1}}$ in the case where the outflow is shaped as a complete cone. 

If the gas emission is solely due to the cocoon, one may expect the emitting structure to be hollow. Nevertheless, the channel maps show that the emission in the region of the radio jet is observed in all channel maps, from blue to redshifted channels, suggesting that the geometry of the outflow is not hollow. On the other hand, one must expect an expansion of the surrounding gas as a jet progress through it, and we cannot disregard the possibility that the observed blue and redshifts may have some contributions from the expansion of a cocoon.

We can now compare the mass outflow rate with the AGN accretion rate, calculated as \citep{peterson97}:

\begin{equation}
\dot{m} = {\frac{L_{bol}}{c^2\,\eta}} \approx 1.8 \times 10^{-3} \left({\frac{L_{44}}{\eta}}\right) \mathrm{\,M_\odot\,yr^{-1}}\, ,
\end{equation}

\noindent
where $L_{bol}$ is the nuclear bolometric luminosity, $c$ is the light speed, $\eta$ is the efficiency of conversion of the rest mass energy of the accreted material into radiation power and $L_{44}$ is bolometric luminosity in units of $10^{44} \,\mathrm{erg\,s^{-1}}$. As in \citet{rifsto11}, we calculated $L_{bol} \approx 100\,L(H\alpha)$, where $L(H\alpha)$ is the nuclear luminosity in H$\alpha$. Integrating H$\alpha$ nuclear flux within our angular resolution (0$\farcs$5), we obtain $F(H\alpha) \approx 2.5 \times 10^{-15} \mathrm{erg\,cm^{-2}\,s^{-1}}$, which leads to an H$\alpha$ luminosity of $L(H\alpha) \approx 4.2 \times 10^{41} \mathrm{erg\,s^{-1}}$, calculated as the previous section. With these values we obtain a bolometric luminosity of $L_{bol} \approx 4.2 \times 10^{43} \mathrm{erg\,s^{-1}}$.

Assuming an efficiency of an optically thick and geometrically thin accretion disk \citep{frank02}, $\eta \approx 0.1$, we derive an accretion rate of $\dot{m} \approx 7.5 \times 10^{-3}\, \mathrm{M_\odot\,yr^{-1}}$, which is just somewhat smaller than the mass outflow rate.  

We can finally use the mass outflow rate to estimate the outflow kinetic power as \citep{holt06}:

\begin{equation}
\dot{E} \approx {\frac{\dot{M}_{out}}{2}} (v_{out}^2 + 3\sigma^2) \, ,
\end{equation}

\noindent

where $\sigma$ is the velocity dispersion. Using an average velocity dispersion measured for H$\beta$, $\sigma \approx 150$\,\kms, $470 < v_{out} < 1100$\,\kms and $ 0.26 < \dot{M}_{out} < 0.32\, \mathrm{M_\odot\, yr^{-1}}$, in which we consider the outflow velocity to be along the radio jet axis, we derive a kinetic power of $ 0.3 < \dot{E} < 1.3 \times 10^{41}\,\mathrm{erg\, s^{-1}}$, which is $ 0.06 - 0.3 \,\%\,L_{bol}$. Considering the case of a cocoon-shaped outflow, in which  $v_{out} \approx 150$\,\kms and $ 0.04 < \dot{M}_{out} < 0.08\, \mathrm{M_\odot\, yr^{-1}}$, we obtain $ 1.4  < \dot{E} < 2.9 \times 10^{39}\,\mathrm{erg\, s^{-1}}$, which is $ 0.003 - 0.007 \,\%\,L_{bol}$.

\section{Conclusions}
\label{conc}

We have measured the gas excitation and kinematics in the inner 2.5\,$\times$\,1.7\,kpc$^2$ of the LINER/Seyfert 1 galaxy Arp 102B from optical spectra obtained with the GMOS integral field spectrograph on the Gemini North telescope with a spatial resolution of 245\,pc at the galaxy. Our main goal was to investigate the nature of the nuclear spiral arms discovered in Paper I. Our measurements and analysis have shown that:
\begin{enumerate}
\item 
the gas emission is enhanced in the nuclear spiral arms, although being observed at lower intensity also outside the arms;

\item
the gas density is higher along the arms when compared to gas outside the arms, showing knots with even larger densities at the northern border of the east arm in the region where the radio jet seems to be deflected;

\item
the gas along the arms shows the lowest excitation, typical of LINERs, with line ratios which are not compatible with ionization by stars, precluding the presence of HII regions in the arms;

\item
the LINER-like ratios, combined with a high velocity dispersion at the base of the arms, support the interaction of the jet with the circum-nuclear gas in the east arm, as well as contribution from shocks to the gas excitation;

\item
the gas centroid velocities over most of the field-of-view are well described by rotation in a disk, but there is a clear spatial correlation between the radio jet, the east spiral arm and a lower centroid velocity, supporting an interaction between the radio jet and circumnuclear gas;

\item
in the channel maps, at the location of the east spiral arm and jet, there are both blueshifts and redshifts, which can be interpreted as due to an outflow occurring close to the plane of the sky, tilted towards us by an small angle of $\approx 1^{\circ}$, so that the front part of the outflow is observed in blueshifts but part of the back part is observed in redshift;

\item
the emitting gas mass calculated in the IFU field is $3.1 \pm 2.7 \times 10^5 M_\odot$;

\item
the estimated mass outflow rate along the east spiral arm is in the range $0.26 - 0.32\, \mathrm{M_\odot\, yr^{-1}}$ (depending on the assumed outflow geometry), which is about two orders of magnitude higher than the AGN mass accretion rate, $7.5 \times 10^{-3}\, \mathrm{M_\odot\,yr^{-1}}$;

\item
we estimate a kinetic power for the outflow of $0.3 - 1.3 \times 10^{41}\,\mathrm{erg\, s^{-1}}$, which correspond to $0.06 - 0.3\,\%$ of the bolometric luminosity.

On the basis of the above characteristics, we propose a scenario in which the gas observed in the inner kiloparsec of Arp 102B was captured in an interaction with its companion, Arp 102A and settled in a rotating disk in the inner kiloparsec. The inflow of this gas towards the nucleus ignited the nuclear activity, giving origin to a radio jet launched close to the plane of the sky. Interaction of the jet with the rotating gas gave origin to the arms which are regions of enhanced emission due to compression by the radio jet. The jet, launched approximately along the line of nodes of the rotating disk, gets deflected to high disk latitudes in a high density gas knot, dragging along gas located in the disk, forming the spiral arms. The enhanced emission in the arms can also be due to additional contribution from shock ionization by the radio jet.

Besides the outflow along the radio jet, we have also found strong evidence for another, more compact ($\sim 300$\,pc) bipolar outflow, whose approaching and receding sides are, respectively, to the south-east and north-west of the nucleus. This is approximately perpendicular to the line of nodes of the large scale disk and remarkably, approximately perpendicular to the radio jet. The observed extended double-peaked H$\alpha$ emission, seen in the PC3 component and in polarized light in a previous study -- can be plausibly attributed to scattering in this outflow and the position angle of polarization is consistent with the outflow being perpendicular to the radio axis. However, we cannot exclude the possibility that this emission perpendicular to the radio jet is due to a smaller-scale rotating disk, which may play a role in fuelling the AGN.

Our observations do not give constraints to the mechanism responsible for the one-sidedness of the radio jet. We conclude that in order to distinguish possible mechanisms it is necessary to have more sensitive radio observations, possibly with JVLA. Higher resolution observations could add constraints to the younger perpendicular outflow.

\end{enumerate}

\section*{Acknowledgments}

We thank the referee for relevant suggestions which helped to improve this paper. This work is based on observations obtained at the Gemini Observatory, which is operated by the Association of Universities for Research in Astronomy, Inc., under a cooperative agreement with the NSF on behalf of the Gemini partnership: the National Science Foundation (United States), the National Research Council (Canada), CONICYT (Chile), the Australian Research Council (Australia), Minist\'{e}rio da Ci\^{e}ncia, Tecnologia e Inova\c{c}\~{a}o (Brazil) and Ministerio de Ciencia, Tecnolog\'{i}a e Innovaci\'{o}n Productiva (Argentina). This work has been partially supported by the Brazilian institution CNPq. This material is based upon work supported in part by the National Science Foundation under Award No. AST-1108786.


\begin{thebibliography}{2}

\bibitem[\protect\citeauthoryear{Allen et al.}{2008}]{allen08} Allen, M.G., Groves, B.A., Dopita, M.A. Sutherland, R.S.; Kewley, L.J., 2008, ApJS, 178, 20

\bibitem[\protect\citeauthoryear{Allington-Smith et al.}{2002}]{allington02} Allington-Smith, J. R., 2002, PASP, 114, 892

\bibitem[\protect\citeauthoryear{Antonucci et al.}{1996}]{antonucci96} Antonucci, R., Hurt, T., Agol, E., 1996, ApJ, 456, 25

\bibitem[\protect\citeauthoryear{Baldwin et al.}{1981}]{baldwin81} Baldwin, J.A., Phillips, M.M., Terlevich, R., 1981, PASP, 93, 5

\bibitem[\protect\citeauthoryear{Barbosa et al.}{2006}]{barbosa06} Barbosa, F.K.B., Storchi-Bergmann, T., Cid Fernandes, R., Winge, C., Schmitt, H., 2006, MNRAS, 371, 170

\bibitem[\protect\citeauthoryear{Bertola et al.}{1991}]{bertola91} Bertola, F., Bettoni, D., Danziger, J., Sadler, E., Sparke, L., de Zeeuw, T., 1991, ApJ, 373, 369

\bibitem[\protect\citeauthoryear{Biermann et al.}{1981}]{biermann81} Biermann, P., Preuss, E., Kronberg, P.P., Schilizzi, R.T., Shaffer, D.B., 1981, ApJ, 250, 49

\bibitem[\protect\citeauthoryear{Caccianiga et al.}{2001}]{caccianiga01} Caccianiga, A., Marchã, M.J.M., Thean, A., Dennett-Thorpe, J., 2001, MNRAS, 328, 867

\bibitem[\protect\citeauthoryear{Cardelli, Clayton \& Mathis}{1989}]{cardelli89} Cardelli, J.A., Clayton, G.C., \& Mathis, J.S., 1989, ApJ, 345, 245

\bibitem[\protect\citeauthoryear{Cid Fernandes et al.}{2009}]{cid09} Cid Fernandes, R., Schoenell, W., Gomes, J.M., Asari, N.V., Schlickmann, M., Mateus, A., Stasinskas, G., Sodré, L., Jr., Torres-Papaqui, J.P., Seagal Collaboration, 2009, RMxAC, 35, 127

\bibitem[\protect\citeauthoryear{Chen, Halpern \& Filippenko}{1989}]{chen89_1} Chen, K., Halpern, J.P., \& Filippenko, A.V., 1989, ApJ, 339, 742

\bibitem[\protect\citeauthoryear{Chen \& Halpern}{1989}]{chen89_2} Chen, K. \& Halpern J.P., 1989, ApJ, 344, 115

\bibitem[\protect\citeauthoryear{Corbett et al.}{1998}]{corbett98} Corbett, E.A., Robinson, A., Axon, D.J., Young, S., Hough, J.H., 1998, MNRAS, 296, 721



\bibitem[\protect\citeauthoryear{Davies et al.}{2009}]{davies09} Davies, R.I., Maciejewski, W., Hicks, E.K.S., Tacconi, L.J., Genzel, R., Engel, H., 2009, ApJ, 702, 114

\bibitem[\protect\citeauthoryear{de Vaucouleurs et al.}{1991}]{devaucouleurs91} de Vaucouleurs, G., de Vaucouleurs, A., Corwin, H.H., Buta, R.J., Paturel, G., Fouque, P., 1991, Third Reference Catalogue of Bright Galaxies (New York: Springer)


\bibitem[\protect\citeauthoryear{Eracleous \& Halpern}{2004}]{eracleous04} Eracleous, M. \& Halpern, J. P., 2004, ApJS, 150, 181

\bibitem[\protect\citeauthoryear{Fathi et al.}{2006}]{fathi06} Fathi, K., Storchi-Bergmann, T., Riffel, R.A., Winge, C., Axon, D.J., Robinson, A., Capetti, A., Marconi, A., 2006, ApJ, 641, L25

\bibitem[\protect\citeauthoryear{Fathi et al.}{2011}]{fathi11} Fathi, K. Axon, D.J., Storchi-Bergmann, T., Kharb, P., Robinson, A., Marconi, A., Maciejewski, W., Capetti, A., 2011, ApJ, 736, 77

\bibitem[\protect\citeauthoryear{Frank, King \& Raine}{2002}]{frank02} Frank, J., King, A., Raine, D.J., 2002, Accretion Power in Astrophysics, 3nd. ed., Cambridge Univ. Press, Cambridge

\bibitem[\protect\citeauthoryear{Flohic \& Eracleous}{2008}]{floera08} Flohic, H.M.L.G., \& Eracleous, M., 2008, ApJS, 622, 178

\bibitem[\protect\citeauthoryear{Groves et al.}{2004a}]{groves04I} Groves, B.A., Dopita, M.A., Sutherland, R.S., 2004, ApJS, 153, 9

\bibitem[\protect\citeauthoryear{Groves et al.}{2004b}]{groves04II} Groves, B.A., Dopita, M.A., Sutherland, R.S., 2004, ApJS, 153, 75

\bibitem[\protect\citeauthoryear{Holt et al.}{2006}]{holt06} Holt, J., Tadhunter, C., Morganti, R., Bellamy, M., González Delgado, R.M., Tzioumis, A., Inskip, K.J., 2006, MNRAS, 370, 1633

\bibitem[\protect\citeauthoryear{Holt et al.}{2008}]{holt08} Holt, J., Tadhunter, C., Morganti, R., 2008, MNRAS, 387, 639


\bibitem[\protect\citeauthoryear{Lucy}{1974}]{lucy74} Lucy, L.B., 1974, AJ, 79, 745

\bibitem[\protect\citeauthoryear{Kewley et al.}{2006}]{kewley06} Kewley, L.J., Groves, B., Kauffmann, G., Heckman, T.,
2006, MNRAS, 372, 961

\bibitem[\protect\citeauthoryear{Kharb et al.}{2006}]{kharb06} Kharb, P., O'Dea, C.P., Baum, S.A., Colbert, E.J.M., Xu, C.,
2006, ApJ, 652, 177

\bibitem[\protect\citeauthoryear{Maciejewski}{2004}]{maciejewski04} Maciejewski, W., 2004, MNRAS, 354, 892

\bibitem[\protect\citeauthoryear{Martini \& Pogge}{1999}]{marpog99} Martini, P., Pogge, R.W., 1999, AJ, 118, 2646

\bibitem[\protect\citeauthoryear{Martini et al.}{2003}]{martini03} Martini, P., Regan, M.W., Mulchaey, J.S., Pogge, R.W., 2003, ApjS, 146, 353

\bibitem[\protect\citeauthoryear{Newman et al.}{1997}]{newman97} Newman, J.A., Eracleous, M., Filippenko, A.V., Halpern, J.P., 1997, ApJ, 485, 570


\bibitem[\protect\citeauthoryear{Osterbrock \& Ferland}{2006}]{ostfer06} Osterbrock, D.E., Ferland, G.J., 1989, Astrophysics of Gaseous Nebulae and Active Galactic Nuclei, 2nd. ed., University Science Books, California

\bibitem[\protect\citeauthoryear{Peterson}{1997}]{peterson97} Peterson, B.M., 1997, An Introduction to Active Galactic Nuclei, Cambridge University Press, Cambridge



\bibitem[\protect\citeauthoryear{Prieto et al.}{2005}]{prieto05} Prieto, M.A., Maciejewski, W., Reunanen, J., 2005, AJ, 130, 1472

\bibitem[\protect\citeauthoryear{Pogge \& Martini}{2002}]{pogmar02} Pogge, R.W., Martini, P., 2002, ApJ, 569, 624

\bibitem[\protect\citeauthoryear{Puschell et al.}{1986}]{puschell86} Puschell, J.J., Moore, R., Cohen, R.D., Owen, F.N., Phillips, A.C., 1986, AJ, 91, 751

\bibitem[\protect\citeauthoryear{Ricci et al.}{2011}]{ricci11} Ricci, T.V., Steiner, J.E., Menezes, R.B., 2011, ApJ, 734, 10

\bibitem[\protect\citeauthoryear{Richardson}{1972}]{richardson72} Richardson, W.H., 1972, JOSA, 62, 55

\bibitem[\protect\citeauthoryear{Riffel et al.}{2008}]{riffel08} Riffel, R.A., Storchi-Bergmann, T., Winge, C., McGregor, P.J., Beck, T., Schmitt, H., 2008, MNRAS, 385, 1129

\bibitem[\protect\citeauthoryear{Riffel \& Storchi-Bergmann}{2011}]{rifsto11} Riffel, R.A., Storchi-Bergmann, T., 2011, MNRAS, 417, 2752

\bibitem[\protect\citeauthoryear{Roy et al.}{2000}]{roy00} Roy, A.L., Ulvestad, J.S., Wilson, A.S., Colbert, E.J.M., Mundell, C.G., Wrobel, J.M., Norris, R.P., Falcke, H., Krichbaum, T., 2000, in Perspectives on Radio Astronomy: Science with Large Antenna Arrays, ed. M. P. van Haarlem, 173

\bibitem[\protect\citeauthoryear{Schnorr M\"uller et al.}{2011}]{schnorr11} Schnorr M\"uller, A., Storchi-Bergmann, T., Riffel, R.A., Ferrari, F., Steiner, J.E., Axon, D.J., Robinson, A., 2011, MNRAS, 413, 149

\bibitem[\protect\citeauthoryear{Sikora et al.}{2007}]{sikora07} Sikora, M., Stawarz, {\L}.; Lasota, J.P., 2007, ApJ, 658, 815

\bibitem[\protect\citeauthoryear{Sim\~oes Lopes et al.}{2007}]{simoes07} Sim\~oes Lopes, R.D., Storchi-Bergmann, T., de Fatima Saraiva, M., Martini, P., 2007, ApJ, 655, 718

\bibitem[\protect\citeauthoryear{Stauffer, Schild \& Keel}{1983}]{stauffer83} Stauffer, J., Schild, R., \& Keel, W.C., 1983, ApJ, 270, 465

\bibitem[\protect\citeauthoryear{Steiner et al.}{2009}]{steiner09} Steiner, J.E., Menezes, R.B., Ricci, T.V., Oliveira, A.S., 2009, MNRAS, 395, 64

\bibitem[\protect\citeauthoryear{Storchi-Bergmann et al.}{1993}]{storchi93} Storchi-Bergmann, T., Baldwin, J. A., Wilson, A. S., 1993, ApJ, 410, L11


\bibitem[\protect\citeauthoryear{Storchi-Bergmann et al.}{2003}]{storchi03} Storchi-Bergmann, T., et al., 2003, ApJ, 598, 956

\bibitem[\protect\citeauthoryear{Storchi-Bergmann et al.}{2007}]{storchi07} Storchi-Bergmann, T., Dors, O. L., Riffel, R. A., 2007, ApJ, 670, 959

\bibitem[\protect\citeauthoryear{Su \& Finkbeiner}{2012}]{su12} Su, M., \& Finkbeiner, D.P., 2012, arXiv:1206.1616

\bibitem[\protect\citeauthoryear{van der Kruit \& Allen}{1978}]{kruit78} van der Kruit, P.C., Allen, R.J., 1978, ARA\&A, 16,
103

\bibitem[\protect\citeauthoryear{Wrobel et al.}{1988}]{wrobel88} Wrobel, J.M., Harrison, B., Pedlar, A., Unger, S.W., 1988, MNRAS, 235,
663


\end{thebibliography}
\end{document}